\documentclass[10pt,conference]{IEEEtran}
\usepackage{amsmath,amssymb,amsfonts}
\usepackage{graphicx}
\usepackage{textcomp}
\usepackage{xcolor}
\usepackage[hyphens]{url}
\usepackage{hyperref}

\usepackage[numbers,sort&compress]{natbib}
\usepackage{amsmath,amsfonts,amsthm}
\usepackage{graphicx}
\usepackage{xcolor}
\usepackage{microtype}
\usepackage[italic]{mathastext}
\usepackage{microtype}
\usepackage{mathastext}
\usepackage{libertine}
\usepackage[T1]{fontenc}
\usepackage{textcomp}
\usepackage{zi4}
\usepackage[all]{nowidow}
\usepackage{flushend}
\usepackage{fancyhdr}
\usepackage{url}
\usepackage{listings}

\usepackage{makecell}

\usepackage{algorithm}
\usepackage{algpseudocode}

\theoremstyle{definition}

\usepackage{newfloat}

\usepackage[small,compact]{titlesec}
\usepackage{caption}
\usepackage{color}
\usepackage[normalem]{ulem}
\usepackage{xspace}
\usepackage{booktabs} %
\usepackage[normalem]{ulem}

\usepackage{makecell}
\usepackage{multirow}
\usepackage{comment}
\usepackage{hhline}
\usepackage[para]{threeparttable} 
\makeatletter
\def\TPT@doparanotes{\par
   \prevdepth\z@ \TPT@hsize
   \TPTnoteSettings
   \parindent\z@ \pretolerance 8
   \linepenalty 200
   \renewcommand\item[1][]{\relax\ifhmode \begingroup
       \unskip
       \advance\hsize 10em %
       \penalty -45 \hskip\z@\@plus\hsize \penalty-19
       \hskip .07\hsize \penalty 9999 \hskip-.15\hsize
       \hskip .01\hsize\@plus-\hsize\@minus.01\hsize 
       \hskip 6em\@plus .3em
      \endgroup\fi
      \tnote{##1}\,\ignorespaces}%
   \let\TPToverlap\relax
   \def\endtablenotes{\par}%
}
\makeatother

\usepackage{hyperref}

\RequirePackage[nameinlink]{cleveref} %
\crefname{chapter}{Chapter}{Chapters}
\crefname{section}{Section}{Sections}
\crefname{subsection}{Subsection}{Subsections}
\crefname{equation}{Equation}{Equations}
\crefname{definition}{Definition}{Definitions}
\crefname{assumption}{Assumption}{Assumptions}
\crefname{theorem}{Theorem}{Theorems}
\crefname{figure}{Figure}{Figures}
\crefname{table}{Table}{Tables}
\crefname{BOX}{Box}{Boxes}
\let\autoref\cref %

\DeclareFloatingEnvironment[
  fileext=lob,
  listname={List of Boxes},
  name=Box,
  placement=htp,
]{BOX}

\newcommand{\revdel}[1]{\ignorespaces}

\newcommand{\squishlist}{
 \begin{list}{$\bullet$}
  { \setlength{\itemsep}{0pt}
     \setlength{\parsep}{3pt}
     \setlength{\topsep}{3pt}
     \setlength{\partopsep}{0pt}
     \setlength{\leftmargin}{1.5em}
     \setlength{\labelwidth}{1em}
     \setlength{\labelsep}{0.5em} } }

\newcommand{\squishlisttwo}{
 \begin{list}{$\bullet$}
  { \setlength{\itemsep}{0pt}
     \setlength{\parsep}{0pt}
    \setlength{\topsep}{0pt}
    \setlength{\partopsep}{0pt}
    \setlength{\leftmargin}{2em}
    \setlength{\labelwidth}{1.5em}
    \setlength{\labelsep}{0.5em} } }

\newcommand{\squishend}{
  \end{list}  }

\newcommand{\betterparagraph}[1]{\noindent \textbf{#1. }}

\crefname{algorithm}{Algorithm}{Algorithms}
\crefname{listing}{Code}{Code}

\definecolor{codegreen}{rgb}{0,0.6,0}
\definecolor{codegray}{rgb}{0.5,0.5,0.5}
\definecolor{codepurple}{rgb}{0.58,0,0.82}
\definecolor{backcolour}{rgb}{0.95,0.95,0.92}

\lstdefinestyle{mystyle}{
    backgroundcolor=\color{backcolour},   
    commentstyle=\color{codegreen},
    keywordstyle=\color{magenta},
    numberstyle=\tiny\color{codegray},
    stringstyle=\color{codepurple},
    breakatwhitespace=false,         
    breaklines=true,                 
    captionpos=b,                    
    keepspaces=true,                 
    numbers=left,                    
    numbersep=5pt,                  
    showspaces=false,                
    showstringspaces=false,
    showtabs=false,                  
    tabsize=2,
    basicstyle=\tiny
}

\newcommand{\accelerator}{\textsc{FIGFLEX}\xspace}

\title{\accelerator: A Mixed-Precision FP-INT GEMM Accelerator Co-Designed with Low-Bit LLM Quantization}

\author{
  \IEEEauthorblockN{Faraz Tahmasebi, Haocheng Xu, Andrew Trauger, Yomal Perera, Zhenyu Tang,
  Nael Chowdhury, and Sitao Huang}
  \IEEEauthorblockA{University of California, Irvine, Irvine, CA, USA \\
  \{tahmasef, haochx5, atrauger, yperera, zhenyut3, naelc, sitaoh\}@uci.edu}
}

\begin{document}
\maketitle

\begin{abstract}

Weight-only quantization keeps LLM activations in floating point and stores weights as low-bit integers, turning the dominant workload into floating-point--integer (FP-INT) general matrix multiplication. Recent quantizers reach a 2--3-bit average with mixed precision, but hardware has not kept pace: GPUs dequantize weights to FP16 before the tensor-core multiply-accumulate, and bit-parallel FP-INT accelerators run at a single weight precision, so fewer bits save memory but not computation.

In this paper, we present \accelerator, an algorithm--hardware co-designed
mixed-precision accelerator whose bit-parallel datapath scales throughput as
precision falls. \accelerator{} contributes
(1)~a shift-add, radix-4 reformulation of FP-INT multiplication that fuses the
multiply-accumulate into a unified, precision-flexible partial-product reduction and
roughly halves the partial products, together with a novel INT3 optimization that
reclaims the idle datapath radix-4 leaves at that precision; and
(2)~a  mixed-precision quantizer
that assigns each channel group its bit-width according to the error GPTQ cannot compensate, with a
per-input-channel column scaling whose power-of-two normalization keeps the scaled
activations out of the FP16 subnormal range, so that the quantized model
maps directly onto \accelerator's datapath with the same accuracy as on a GPU.
On LLaMA-2-7B, \accelerator{} attains $5.85$ and $8.77$ WikiText-2 perplexity at
3- and 2-bit average precision, lower than GPTQ, AWQ, SliM-LLM, AMQ, and QuIP\# (without fine-tuning), while delivering $1.56\times$ and $1.42\times$ higher whole-chip throughput
per area on the linear GEMMs of a 3-bit mixed-precision model than AxCore~\cite{axcore} and FIGLUT~\cite{figlut},
respectively.
\end{abstract}

\section{Introduction}
\label{sec:intro}

Large language models (LLMs) have become foundational to modern natural language processing, and their capability scales with size: state-of-the-art models hold tens to hundreds of billions of parameters, and a 175B-parameter model alone occupies roughly 350\,GB in FP16, which is far beyond the memory of a single accelerator~\cite{figlut,axcore,opt}.
In inference, the linear layers of the transformer dominate both the parameter count and the runtime, and their core operation is general matrix multiplication (GEMM).
Executed in floating point (FP16/BF16), these GEMMs are both memory- and compute-bound and are the primary obstacle to efficient LLM serving.

Weight-only quantization has become the accuracy-preserving standard for compressing LLMs.
Because weights vastly outweigh activations in memory while activations carry accuracy-critical outliers, compressing weights to low-bit integers and keeping activations in floating point preserves quality better than quantizing both~\cite{gptq,awq}. This turns the dominant workload into a
floating-point--integer (\mbox{FP-INT}) GEMM between FP16 activations and low-bit
integer weights.

While INT4 is now an established quality-preserving
precision~\cite{gptq,spinquant,awq}, the frontier has moved lower. Uniform methods
push weights to 2--3 bits with incoherence processing and lattice
codebooks~\cite{quip,aqlm,quipsharp}, and \emph{mixed-precision} methods go further, spending
4 bits only on the small fraction of salient weight groups and 1--2 bits elsewhere
to reach a low \emph{average} precision at little quality cost~\cite{slimllm,amq}.
Mixed precision in particular produces a \emph{per-group bit budget}: the number of
bits each weight group actually needs, which varies across the model. This heterogeneity should translate not only into a cheaper memory footprint---which efficient low-bit weight-storage kernels already realize~\cite{marlin,lutgemm,anyprec,flute}---but also into cheaper computation, turning that compression into a real speedup of the linear GEMMs that dominate LLM runtime.

On the compute side, however, that budget goes unspent. General-purpose GPUs have no
native FP16$\times$INT datapath, so they \emph{dequantize the weights back to FP16}
before the multiply-accumulate and run it on floating-point units~\cite{gptq,awq}; fewer
bits therefore cut memory and bandwidth but not computation. State-of-the-art dedicated FP-INT accelerators remove this dequantization by
\emph{pre-aligning} activations into integer mantissas~\cite{ifpu}, but their efficiency is realized for a fixed precision. FIGNA~\cite{figna} commits to a single bit-width (INT4) and is
inefficient below it or for mixed precision. AxCore~\cite{axcore} fixes the weight quantization to the FP4 format across its E/M variants, and approximates the
multiplier with a logarithmic-domain addition and an error-compensation term, while keeping
the accumulation in floating point---a far larger unit than integer
accumulation. For both, a weight
in fewer bits costs the same to compute, so lowering the weight bit-width does not raise throughput.
Bit-serial architectures have tried to resolve this inefficiency~\cite{stripes,bitpragmatic,ifpu}, and FIGLUT~\cite{figlut} is the state of the art among them: it supports
mixed precision on a single substrate. It splits each weight into binary planes and reads a
look-up table of pre-summed activations selected by the weight pattern. However, its work grows with the weight
bit-width and the table must precompute every signed combination of its $\mu$ grouped
activations (exponential in $\mu$), requiring a large, many-ported flip-flop LUT. %

In this paper, we present \accelerator, an algorithm--hardware co-designed accelerator whose bit-parallel datapath raises throughput as precision falls. Contributions are summarized as:

\begin{itemize}
  \item We identify and quantify the gap left by fixed-precision, bit-parallel FP-INT accelerators: their compute cost does not fall as weight bit-width falls, so they cannot exploit the mixed precision that low-bit LLMs now rely on.

  \item We design a \emph{bit-parallel, precision-flexible} FP-INT processing element built on a radix-4 weight recoding.  A dedicated INT3 optimization fuses the two high bits of a weight pair into one shared lane, recovering the throughput a naive radix-4 wastes on odd precision. The element adds no error of its own (the radix-4 recoding is an exact regrouping and the integer accumulation is exact) with exponent alignment, multiple generation, and normalization hoisted to a shared array edge.
  \item We co-design a  mixed-precision quantizer with a
    dequantization-free epilogue. It assigns each channel group a unique bit-width based on the error
    that GPTQ's compensation leaves behind, rather than by a salience score computed
    before quantization, and it normalizes each input channel with a column scale that is
    rescaled by a power of two ($2^{s}$) so that the scaled activations never fall into FP16 subnormal range the hardware flushes. The quantizer maps directly onto the
    hardware's precision scaling.

\end{itemize}

On a 3-bit mixed-precision model, \accelerator{} delivers $1.56\times$ and $1.42\times$ the whole-chip throughput per area of AxCore~\cite{axcore}, the state-of-the-art approximate FP-INT engine, and FIGLUT~\cite{figlut}, the state-of-the-art precision-scalable one: unlike AxCore, its throughput follows the weight bit-width, and unlike FIGLUT, it does so bit-parallel. The radix-4 recoding is exact, and with the shared-exponent alignment of the RTL the quantized models reach the same perplexity on \accelerator{} as on a GPU. On LLaMA-2-7B, \accelerator{} attains 5.85 and 8.77 WikiText-2 perplexity at 3- and 2-bit average precision, lower than state-of-the-art low-bit post-training methods (GPTQ~\cite{gptq}, AWQ~\cite{awq},  AMQ~\cite{amq}, SliM-LLM~\cite{slimllm}, and QuIP\#~\cite{quipsharp} without fine-tuning); QuIP\#, whose weights an FP-INT array cannot consume directly, keeps an edge in C4 perplexity and zero-shot accuracy at 2 bits.

\section{Background and Motivation}
\label{sec:bg}

\subsection{LLM Computation}
\label{sec:bg:compute}

An LLM is a stack of decoder blocks with two types of GEMMs: the linear layers (Q/K/V and output projections, feed-forward) multiply activations by static weights, while attention multiplies activations by activations. For hidden size $d$ and context length $l$, a block's linear GEMMs cost $\mathcal{O}(d^2)$ MACs per token against attention's $\mathcal{O}(ld)$, so the linear GEMMs dominate compute, and only they involve the quantized integer weights.

{Figure~\ref{fig:motivation} decomposes a full request of LLaMA-3.1-405B~\cite{llama3} (3-bit weights and KV cache, H200-class GPU, batch $128$, vllm~\cite{vllm}) into linear GEMM, attention, and memory. The linear GEMM takes 71--85\% of the request time: almost all of prefill, and a substantial share of decode, where streaming the weights and the KV cache becomes equally critical. Lowering the GPU's weights and KV cache to 2 bits shortens a request by only 2--9\%, since the dequantized GEMM still runs at the FP16 rate. This work targets accelerating the linear GEMMs. Our evaluation measures these GEMM layers; the \accelerator bars of Figure~\ref{fig:motivation} are a first-order end-to-end estimate built on the measured GEMM rates, with the rest of the request modeled (Section~\ref{sec:hw}).

\begin{figure}[t]
\centering
\includegraphics[width=\linewidth]{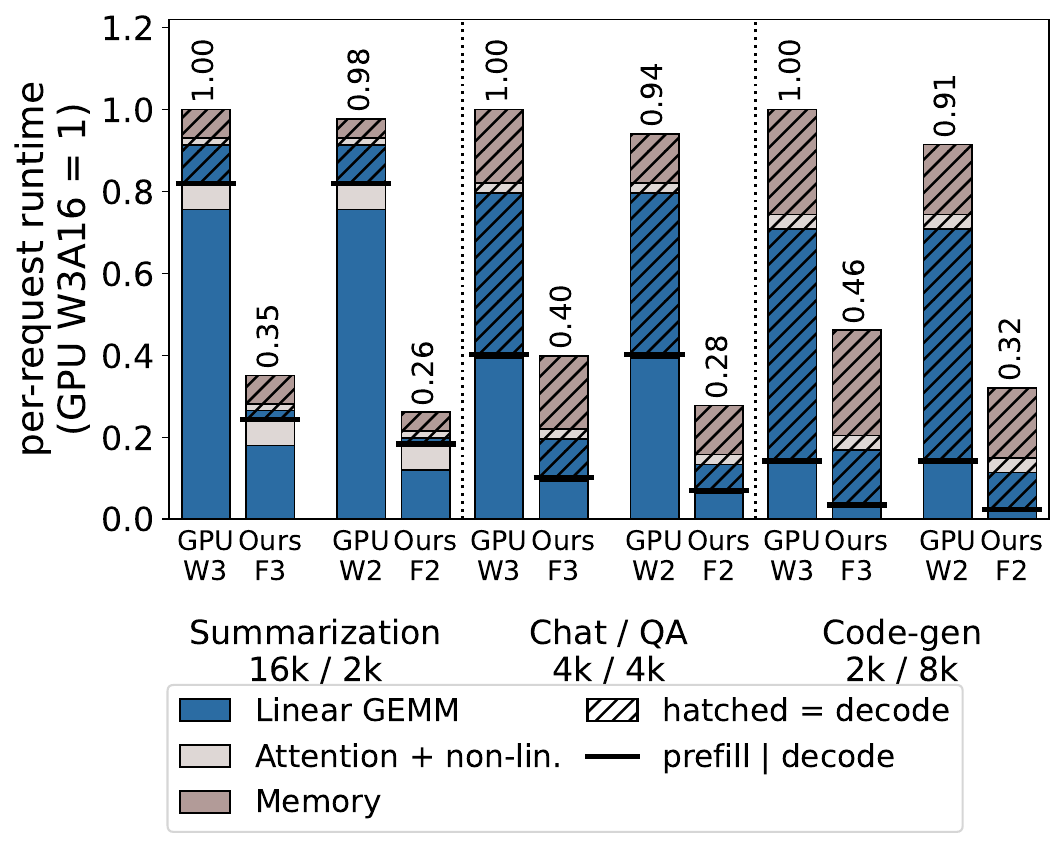}
\vspace{-7mm}
\caption{\small \textbf{The linear GEMM (blue bars) share of the total per-request runtime of LLaMA-3.1-405B in various tasks.}
The quantized model (GQA, weights and KV cache at 3 or 2 bits) is deployed on an
H200-class GPU; each bar is split into linear GEMM, attention, and memory; solid segments are
prefill, hatched are decode. GPU low-bit kernels (W3A16, W2A16; GPU W3A16 = 1 per task) are
compared with \accelerator at the same bit-width (Ours F3, F2), placed in the GPU's tensor-core compute
area at the same bandwidth: its GEMMs run at the measured compute density
(Figure~\ref{fig:tops_per_area}) and F3/F2 rates (Table~\ref{tab:fractions}), and attention and
memory are modeled as for the GPU (first-order model).
}
\vspace{-3mm}
\label{fig:motivation}
\end{figure}

\subsection{Weight-Only Quantization}
\label{sec:bg:woq}
Post-training quantization (PTQ) compresses a trained model without retraining. Weight-only PTQ stores the weights as low-bit integers and keeps the activations in FP16. Weights are quantized per group of 128 input channels with an asymmetric affine code: a $p$-bit weight is stored as $q=\mathrm{clip}(\lfloor w/s+z\rceil,0,2^p-1)$ and decoded as $\hat{w}=s(q-z)$, with one scale $s$ and one integer zero-point $z$ per group.
 
GPTQ~\cite{gptq}, the basis of our quantizer, minimizes the layer-wise output error $\lVert X(W-\hat{W})^\top\rVert_F^2$ with the calibration Hessian $H=X^\top X$. It quantizes the columns one at a time, most salient first (act-order), and propagates each column's rounding error to the columns not yet quantized through $H^{-1}$. Those columns compensate part of the error, and only the remainder reaches the layer output. AWQ~\cite{awq}, OmniQuant~\cite{omniquant}, and rotation-based methods~\cite{quipsharp,quarot,spinquant} reach near-FP16 quality at 4 bits by other means.
 
Below 4 bits, \emph{uniform} methods rely on incoherence processing and lattice, additive, vector, or trellis codebooks~\cite{quip,quipsharp,aqlm,vptq,qtip}, whereas \emph{mixed-precision} methods vary the bit-width across weight groups: SliM-LLM~\cite{slimllm} assigns widths by a salience metric and AMQ~\cite{amq} searches a per-layer allocation. The resulting per-group, non-power-of-two bit-widths are what fixed-precision hardware cannot exploit.

\subsection{FP-INT GEMM Computation}
\label{sec:bg:fpint}

GPUs have no FP16$\times$INT datapath. Low-bit kernels such as CUTLASS mixed-input~\cite{cutlass}, Marlin~\cite{marlin}, and FLUTE~\cite{flute} \emph{dequantize} the weights to FP16 on the fly and issue ordinary tensor-core MACs, so their arithmetic cost is independent of the bit-width and their gain saturates once the GEMM is compute-bound (prefill, or decode at moderate batch). LUT-GEMM~\cite{lutgemm} and Any-Precision LLM~\cite{anyprec} process binary weight planes through lookup tables on the CUDA cores, which helps memory-bound decode but is slower than the tensor-core kernels once compute dominates; the latter are our GPU baseline (Figure~\ref{fig:motivation}).

\textbf{FIGNA}~\cite{figna} is a bit-parallel FP16$\times$INT4 unit: a block of activations is \emph{pre-aligned} to its maximum exponent~\cite{ifpu}, multiplied and accumulated against the integer weights in integer hardware, and converted back to floating point once. \textbf{AxCore}~\cite{axcore} is multiplier-free but \emph{approximate}: it replaces the multiply with a logarithmic-domain addition and an error-compensation term, accumulates in a per-PE partial floating-point adder whose normalization is deferred to a shared module, and commits to the FP4 format. Both process every weight at one width, so a weight stored in fewer bits is no cheaper to compute, and lowering the weight bit-width does not raise their throughput.

\textbf{Bit-serial} designs make throughput scale with precision by processing one weight bit-plane per step~\cite{stripes,bitpragmatic}. \textbf{FIGLUT}~\cite{figlut}, the state of the art, performs $\mu$ MACs per bit-plane with one read of a table that holds all $2^\mu$ signed sums of $\mu$ grouped activations. The table and its selection logic grow exponentially in $\mu$ and dominate the PE, and a $b$-bit weight still takes $b$ reads.

Each of these designs trades one property for the other: GPUs gain only memory
bandwidth, FIGNA and AxCore are efficient but fixed-precision, and FIGLUT parallelizes its
bit-serial computation only at the cost of an exponentially larger datapath. We
co-design a \emph{bit-parallel} FP-INT datapath whose throughput scales with the weight bit-width.

\section{\accelerator Architecture}
\label{sec:arch}

This section presents \accelerator, a bit-parallel FP-INT GEMM accelerator.  Its processing element has no multiplier: each weight is recoded
into radix-4 digits, each digit selects a precomputed multiple of the activation through a $4{:}1$
multiplexer, and one integer adder tree forms the products and their sum at once
(Sections~\ref{sec:arch:principles} and~\ref{sec:arch:radix4}). The same PE therefore runs INT4,
INT3, INT2, and INT1 weights at 8, 10.67, 16, and 32 MAC/cycle. A high-bit lane fusion removes the
half lane that radix-4 leaves idle at INT3 (Section~\ref{sec:arch:int3}), and
Section~\ref{sec:arch:accel} presents the full accelerator and its mixed-precision dataflow.

\subsection{Design Principles}
\label{sec:arch:principles}

\betterparagraph{Compute in the integer domain}
A GPU dequantizes low-bit weights to FP16 before its tensor cores multiply, so compression saves memory but not
arithmetic (Section~\ref{sec:bg:fpint}). \accelerator{} keeps the dot product in the integer
domain: as in integer-domain FP-INT units~\cite{ifpu,figna}, a block of activations is pre-aligned
to its maximum exponent with $p{+}2$ guard bits for $p$-bit weights, so the dot product becomes an
integer reduction that is converted back to floating point once.

\betterparagraph{Fusing multiplication and accumulation}
After pre-alignment, multiplying a significand by an integer weight and accumulating the products
are the same primitive: a sum of shifted significands. \accelerator{} performs both in one binary
adder tree, with neither a multiplier per product nor a bit-serial look-up~\cite{figlut}. This
fusion is what makes the datapath precision-flexible: a weight with fewer bits contributes fewer
shifted terms, so it occupies fewer adder-tree inputs, and the freed inputs serve more weights in
the same cycle.

\begin{figure}[t]
\centering
\includegraphics[width=\linewidth]{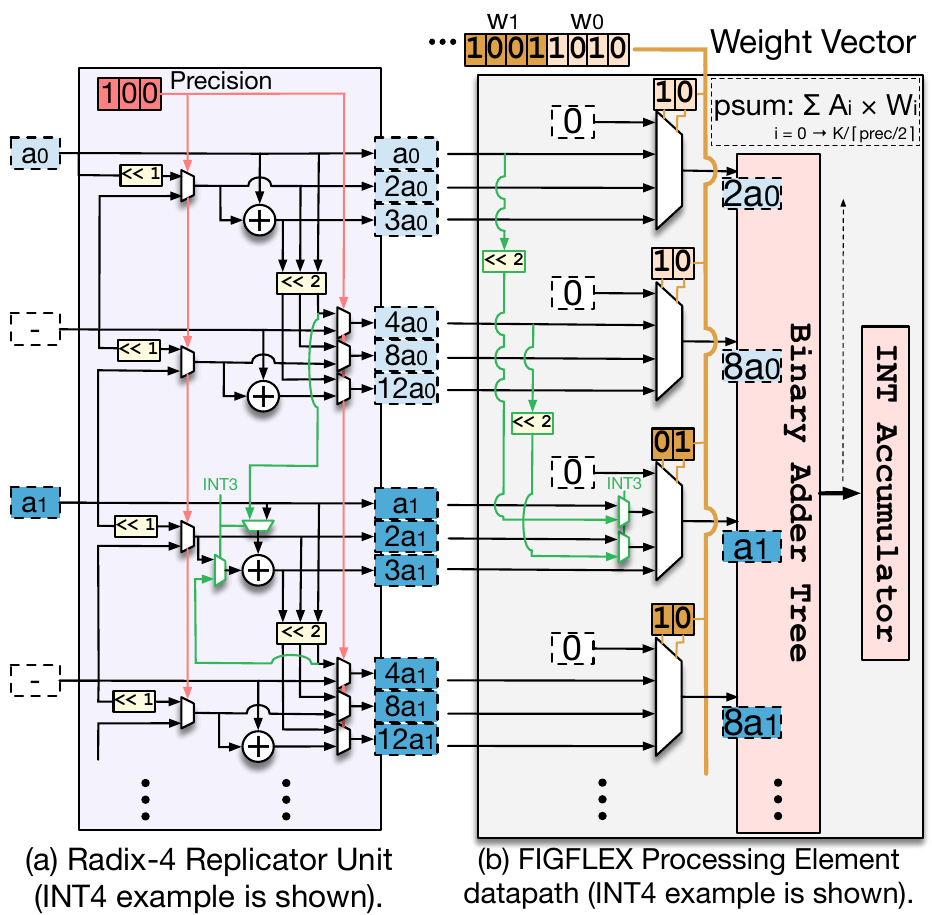}
\vspace{-7mm}
\caption{\small Radix-4 datapath. \emph{(a)} Per-row replicator. \emph{(b)} PE: per-lane
$4{:}1$ multiplexers feed one adder tree and an integer accumulator. INT3 fused-lane wiring
(Section~\ref{sec:arch:int3}) in green.}
\vspace{-3mm}
\label{fig:pe_design}
\end{figure}

\subsection{Radix-4 FP-INT Datapath}
\label{sec:arch:radix4}

While our architecture can be constructed for any activation precision, we elaborate on
the design for FP16$\times$INT arithmetic. The core operation of weight-only quantized
inference is a dot product between an FP16
activation vector and an integer weight vector,
\begin{equation}
\label{eq:dot}
y=\sum_{k} a_k\,w_k,\qquad
a_k=(-1)^{s_k} m_k\,2^{\,e_k-\beta},\qquad
w_k=\sum_{i=0}^{P-1} b_{k,i}\,2^{i},
\end{equation}
where $m_k$ is the 11-bit activation significand (with implicit one), $e_k$ its
exponent, $\beta$ a fixed bias, and $b_{k,i}$ the $i$-th bit of the $P$-bit weight
(sign, scale, and zero-point are handled outside the PE-array datapath,
Section~\ref{sec:arch:accel}). The pre-aligner replaces the factor $2^{\,e_k-\beta}$ by
a right shift of each significand to the shared exponent $E=\max_k e_k$, giving aligned
integer significands $\hat m_k = m_k \gg (E-e_k)$, so the dot product is an integer
reduction over $\hat m_k w_k$.

\betterparagraph{Radix-4 recoding}
Grouping the weight bits in pairs turns a $P$-bit weight into $\lceil P/2\rceil$
radix-4 digits,
\begin{equation}
\label{eq:radix4}
w_k=\sum_{j=0}^{\lceil P/2\rceil-1} d_{k,j}\,4^{\,j},\qquad d_{k,j}\in\{0,1,2,3\},
\end{equation}
and each digit contributes a single partial product $d_{k,j}\,\hat m_k$ instead of one
per bit. The partial product is therefore one of four multiples of the aligned
significand, $\{0,\,\hat m_k,\,2\hat m_k,\,3\hat m_k\}$, and the digit's place value
$4^{j}$ is a fixed left shift of $2j$ applied at the edge. Radix-4 is the sweet spot: radix-2 needs
$1.29\times$ the area per MAC, and radix-8/16 spend more on precomputed operands than they save in
the adder tree (\autoref{tab:radix}).

\betterparagraph{Processing element}
\accelerator{} realizes this on a $K$-lane processing element (PE)
(Figure~\ref{fig:pe_design}b shows 4 lanes of the PE). The weight vector is loaded from the
weight aligner atop the PE array and distributed among the lanes. Each lane then uses its
2-bit weight digit to drive a $4{:}1$ multiplexer that selects the corresponding multiple.
Of the four multiples, three cost no logic: $\hat m_k$ and
$2\hat m_k=\hat m_k\!\ll\!1$ are free wires and $0$ is trivial; only $3\hat m_k$ is a
\emph{hard} multiple requiring an addition $\hat m_k+2\hat m_k$.  {Because it
depends only on the activation, it is formed once per row at the edge.} A PE is thus a
bank of $K$ selects feeding one $K$-input binary adder tree and an integer accumulator;
it contains no multiplier and no shifter.

\betterparagraph{Radix-4 replicator}
The shift networks are hoisted to a per-row \emph{radix-4 replicator}
(Figure~\ref{fig:pe_design}a). For each activation it receives the signed significand
$\pm\hat m_k$ from the exponent aligner, forms its hard multiple $\pm 3\hat m_k$---folding
the activation sign in, so the PE needs no negation---applies each lane's radix-4 position
shift, and broadcasts the $K$ lane operands to every PE in the row. Note that the datapath
is designed for non-negative weight codes, as produced by asymmetric quantization; it can
support symmetric quantization by simply using $2^{P-1}$ as the zero point ($P$ is the
weight precision). The two shift networks (exponent alignment and multiple/position
generation), which would dominate a per-PE implementation, are thus paid once per row.

\betterparagraph{Precision scaling}
Because each lane carries two weight bits, the $K$-lane PE consumes $2K$ weight
bits per cycle, so a $b$-bit weight sustains up to $2K/b$ MAC/cycle. INT4 spends
$\lceil4/2\rceil{=}2$ lanes per weight (for our choice $K{=}16$, \textbf{8} MAC/cycle), INT2 one lane per weight (\textbf{16}), and INT1 two weights per lane
(\textbf{32}), the lane's two digit bits selecting $\{0,a_i,a_{i+1},a_i{+}a_{i+1}\}$; reconfiguring precision changes only the replicator's routing,
while the select lanes, adder tree, and accumulator are shared. INT3 is the exception:
$\lceil3/2\rceil{=}2$ lanes per weight, but the high digit is a single bit occupying a
two-bit lane, so half of every second lane is idle and the naive rate is only $8$
MAC/cycle---no better than INT4. Recovering that waste is the subject of
Section~\ref{sec:arch:int3}.

\begin{table}[b]
\centering
\caption{{Per-lane radix-4 operation: every lane is the same $4{:}1$ mux, and only the
replicator's routing changes with precision. $a_i$: aligned activations; $\times4$ shifts are wiring.}}
\label{tab:lanes}
\resizebox{\columnwidth}{!}{%
\begin{tabular}{c cc cc}
\toprule
& \multicolumn{2}{c}{\textbf{INT4} (2 lanes/wt)}
& \multicolumn{2}{c}{\textbf{INT3} (3 lanes/2\,wt)}\\
\cmidrule(lr){2-3}\cmidrule(lr){4-5}
Lane & mux inputs & digit & mux inputs & digit\\
\midrule
1 & $\{0,a_0,2a_0,3a_0\}$    & $w_{0[1:0]}$ & $\{0,a_0,2a_0,3a_0\}$ & $w_{0[1:0]}$\\
2 & $\{0,4a_0,8a_0,12a_0\}$  & $w_{0[3:2]}$ & $\{0,a_1,2a_1,3a_1\}$ & $w_{1[1:0]}$\\
3 & $\{0,a_1,2a_1,3a_1\}$    & $w_{1[1:0]}$ & $\{0,4a_0,4a_1,4(a_0{+}a_1)\}^{\dagger}$ & $\{w_{1[2]},w_{0[2]}\}$\\
4 & $\{0,4a_1,8a_1,12a_1\}$  & $w_{1[3:2]}$ & $\{0,a_2,2a_2,3a_2\}$ & $w_{2[1:0]}$\\
5 & $\{0,a_2,2a_2,3a_2\}$    & $w_{2[1:0]}$ & $\{0,a_3,2a_3,3a_3\}$ & $w_{3[1:0]}$\\
6 & $\{0,4a_2,8a_2,12a_2\}$  & $w_{2[3:2]}$ & $\{0,4a_2,4a_3,4(a_2{+}a_3)\}^{\dagger}$ & $\{w_{3[2]},w_{2[2]}\}$\\
\midrule
\multicolumn{5}{@{}l}{\textbf{INT2} (1 lane/wt): lane $i=\{0,a_i,2a_i,3a_i\}$, digit $w_{i[1:0]}$}\\
\multicolumn{5}{@{}l}{\textbf{INT1} (2 wt/lane): lane $i=\{0,a_{2i},a_{2i+1},a_{2i}{+}a_{2i+1}\}$, digit $\{w_{2i+1},w_{2i}\}$}\\
\bottomrule
\end{tabular}}

\vspace{2pt}
{\scriptsize $^{\dagger}${Fused high-bit lane (Section~\ref{sec:arch:int3}).}}
\end{table}

\subsection{INT3 Optimization}
\label{sec:arch:int3}

{Our novel \emph{high-bit lane fusion} reclaims the idle half lane by} processing INT3 weights \emph{in pairs} and \emph{fusing} their
two high bits into one shared lane.

\betterparagraph{Fused high-bit lane}
Take a weight pair $(w_0,w_1)$ with aligned significands $a_0,a_1$. Their low digits
$w_{0[1:0]},w_{1[1:0]}$ occupy two ordinary radix-4 lanes over
$\{0,a_i,2a_i,3a_i\}$. The two high bits contribute $4\,w_{0[2]}a_0+4\,w_{1[2]}a_1$.
If the corresponding weight bits are aligned and paired, the PE can use a single $4{:}1$
multiplexer to produce this contribution: fed the inputs
$\{0,\,4a_0,\,4a_1,\,4(a_0{+}a_1)\}$ and the 2-bit code $\{w_{1[2]},w_{0[2]}\}$, it
realizes
\begin{equation}
\label{eq:int3}
4\,w_{0[2]}a_0+4\,w_{1[2]}a_1=
\begin{cases}
0 & 00\\ 4a_0 & 01\\ 4a_1 & 10\\ 4(a_0{+}a_1) & 11.
\end{cases}
\end{equation}

With high-bit lane fusion, two weights need only three lanes instead of four---two low
digits plus one fused high digit---so INT3 packs the array with no idle bits and sustains
$2K/3$ MAC/cycle: a frame of 32 weights (16 pairs)
fills the $3K{=}48$ lanes of three consecutive cycles, so no lane is left idle even though $3\nmid K$
(10.67 MAC/cycle for $K{=}16$), up from $K/2$.
The fused lane reuses the same $4{:}1$ mux as every other lane; only its \emph{inputs}
change. The replicator forms $4(a_0{+}a_1)$ on the second output using the lane's existing
adder (the $\times4$ is wiring) and leaves the first output empty. The PE taps the required
$4a_0,4a_1$ from the two preceding lanes through two $2{:}1$ muxes.  \autoref{fig:pe_design} shows the INT3
high-bit lane fusion datapath in green.

\betterparagraph{Weight aligner}
Because the fused lane is driven by the two high bits of a pair, the incoming INT3
weight vector is repacked at the array edge by a small \emph{weight aligner}
(Figure~\ref{fig:weight_aligner}): for each pair it emits the two low digits followed by
the two MSBs $\{w_{1[2]},w_{0[2]}\}$ concatenated at the end, so the PE consumes the
three lane selectors contiguously and every third lane receives the fused code. Like the
recoding, the transformation is exact---it merely reorders bits. For INT1, two
consecutive 1-bit weights share one lane; their natural packing already places them on that
lane, so INT1 needs no repacking and runs at twice the INT2 rate.

\begin{figure}[t]
\centering
\includegraphics[width=\linewidth]{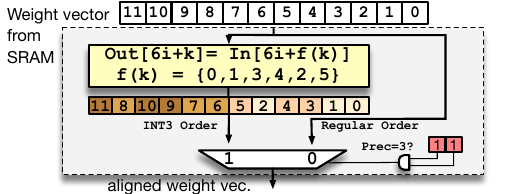}
\vspace{-7mm}
\caption{\small Weight aligner at the array edge. If the precision is INT3,
the MSBs of each weight pair $\{w_{(i+1)[2]},w_{i[2]}\}$ are concatenated at the end.}
\vspace{-3mm}
\label{fig:weight_aligner}
\end{figure}

Table~\ref{tab:lanes} summarizes the per-lane operation across precisions.

\subsection{Full Accelerator}
\label{sec:arch:accel}

{Figure~\ref{fig:figflex_accelerator} shows the complete accelerator: an $M\!\times\!N$ array of
radix-4 PEs, each row served by three shared edge units---the exponent aligner, the radix-4
replicator (Section~\ref{sec:arch:radix4}), and the weight aligner (Section~\ref{sec:arch:int3})---so
the array cost is dominated by the small mux-plus-adder-tree PEs. Each PE column drains its integer
accumulator through a shared INT$\rightarrow$FP converter that restores the block exponent, and a
zero-point unit forms the per-group activation sums used by the epilogue.}

\betterparagraph{Mixed-precision dataflow}

\autoref{fig:Layer_Compute} shows how a layer is processed, including how the column scales
are absorbed into the activations. Our quantizer (Section~\ref{sec:algo}) assigns precisions per
output channel. Offline, the
channels of each 128-input-channel group are sorted by precision, which splits
a layer into three uniform-precision GEMMs ($W_{x+1}$, $W_x$, $W_{x-1}$) that any dataflow can run;
the stored precision tags identify each partial sum's channel, so outputs land in natural order
without an index table. Within a group the array is output-stationary; across groups the loop order is
input-stationary: a group's activations stay in the row buffers while the output tiles of all three
precision parts stream through, so they are fetched once. A 128-channel tile that straddles two precisions runs in the higher mode
(lower-precision codes zero-pad), which keeps utilization above 99\% on Llama-3.1-8B
(Table~\ref{tab:fractions}) and above 93\% for its smallest layers. Tokens map to the 16 PE rows,
so a batch of 16 or more fills the array; smaller decode batches are memory-bound
(Figure~\ref{fig:motivation}) and gain from the reduced weight traffic. Every mode reads one 32-bit
weight word per column per cycle, the act-order permutation is a static gather on the activation
read path, and the epilogue drains each group while the next computes, even at the INT1 rate.
\begin{figure}[t]
\centering
\includegraphics[width=\linewidth]{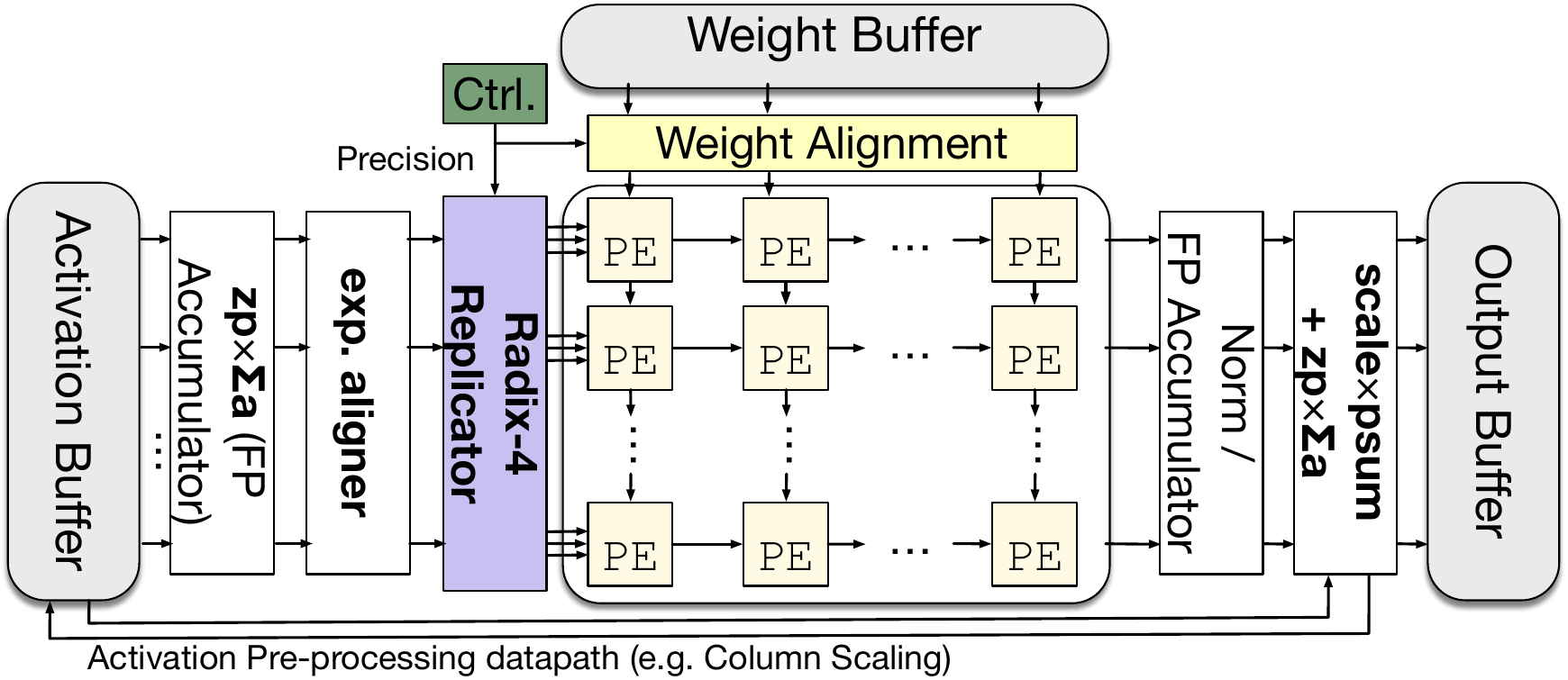}
\vspace{-7mm}
\caption{\small {\accelerator{} overview: shared per-row edge units feed an
$M\!\times\!N$ array of radix-4 PEs; per-column converters and the epilogue produce the FP output.}}
\vspace{-3mm}
\label{fig:figflex_accelerator}
\end{figure}

\begin{figure}[t]
\centering
\includegraphics[width=\linewidth]{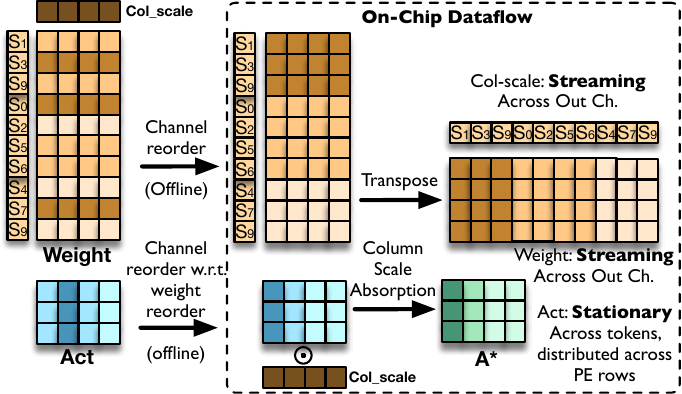}
\vspace{-7mm}
\caption{\small Dataflow example. Offline, the weight channels are reordered by precision and the activation
channels follow the same order. On chip, the column scales are absorbed into the activations ($A^*$),
which stay stationary across the PE rows while the weights stream across output channels.}
\vspace{-3mm}
\label{fig:Layer_Compute}
\end{figure}

\betterparagraph{Dequantization-free epilogue}
Sign, per-channel scale, and the per-group affine are applied to the \emph{scalar}
accumulator output, never to the weights, so no dequantization datapath is required.
Folding the per-input-channel column scale $c_k$ into the activation at the pre-aligner
($a'_k=a_k c_k$), a PE emits the integer dot $S_q=\sum_k a'_k q_{k}$ over a group, and
the epilogue computes
\begin{equation}
\label{eq:epilogue}
Y[m,n]=\sum_{g}\text{scale}_{n,g}\Big(S_q-\text{zp}_{n,g}\!\!\sum_{k\in g}\! a'_k\Big),
\end{equation}
where $\sum_{k\in g} a'_k$ is formed once per group and reused across output channels.
No weight is ever reconstructed to FP16; the epilogue is a small pipelined FP unit that
fires once per quantization group.

\section{Proposed Mixed-Precision Quantization}
\label{sec:algo}

\begin{figure*}[h]
\centering
\includegraphics[width=\textwidth]{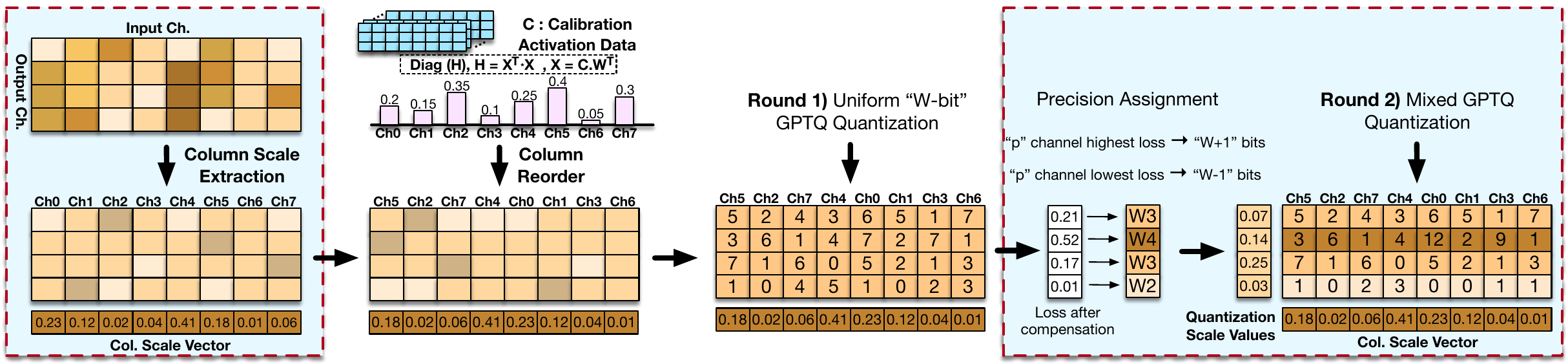}
\vspace{-7mm}
\caption{\small \accelerator's Mixed-Precision Quantization Framework. The \accelerator{} quantizer's contributions are highlighted by dashed boxes.}
\vspace{-4mm}
\label{fig:MPQ_Algorithm}
\end{figure*}

\subsection{Overview}
\label{sec:algo:overview}

\label{sec:algo:prior}
\label{sec:algo:calib}
We build on GPTQ~\cite{gptq} (Section~\ref{sec:bg:woq}), which quantizes every group of a
layer at the same bit-width. Our quantizer (\autoref{fig:MPQ_Algorithm}) adds two contributions:
\emph{(i)}~\textbf{column scaling}, a per-input-channel normalization (\autoref{sec:algo:colscale}),
and \emph{(ii)}~a \textbf{compensation-aware (CA) mixed-precision} assignment over
$\{\text{W4},\text{W3},\text{W2},\text{W1}\}$ per output channel (\autoref{sec:algo:alloc}). Everything
else follows prior work. GPTQ runs with act-order (groups are formed in that order) and $1\%$ damping, block by block on the outputs
of the already-quantized blocks (calibration in \autoref{sec:setup}). A group is $128$ input
channels of one output channel with an FP16 scale and an integer zero point~\cite{gptq,slimllm,awq},
and a channel's precision is set separately in each of its groups. Scales and zero points are fitted by the clip-range search of SliM-LLM~\cite{slimllm}. W1 groups use SliM-LLM's mean-centred
binarization $\mu\pm\alpha$, which is affine in the 1-bit code ($\text{scale}=2\alpha$). Every
precision therefore reduces to $W_{n,k}=c_k\,(q_{n,k}-\text{zp}_{n,g})\,\text{scale}_{n,g}$, the
form the epilogue consumes without dequantization (Eq.~\eqref{eq:epilogue}).

\subsection{Column Scaling}
\label{sec:algo:colscale}

For each input channel $k$ we compute
\begin{equation}
c_k=\tfrac{1}{N}\textstyle\sum_{n}\lvert W_{n,k}\rvert,
\end{equation}
normalize $\widetilde{W}_{n,k}=W_{n,k}/c_k$ before quantization, and fold $c_k$ back at
inference. Because $c_k$ is per input channel, it is equivalently a \emph{per-channel
activation scale} $a'_k=a_k c_k$ and costs nothing in the weight path
(Section~\ref{sec:arch:accel}). Figure~\ref{fig:Layer_Compute} shows that the column scales
can be absorbed into the activations as scalar values prior to the linear GEMM.
The multiply happens once per activation element at the pre-aligner and is shared by all
$N$ output channels ($1/N$ of the MACs). Because $c_k$ is small, $a_k c_k$ can underflow into
the FP16 subnormal range, which the pre-aligner flushes to zero. We therefore rescale $c$ per
layer by a power of two, $c'=c/2^{s}$, and multiply the group scales by $2^{s}$, which leaves
the quantized model unchanged; $s$ places the largest calibration product $|a_k c'_k|$ at
$2^{12}$.
Column scaling flattens the per-channel dynamic range and
is the main reason the aggressive W1 groups remain usable: on Llama-3.1-8B at 2 bits,
removing it raises perplexity from 13.77 to 16.48 (\autoref{tab:ablation}). The scaling runs in the opposite direction to AWQ~\cite{awq}. AWQ multiplies the weight
columns of activation-salient channels by $s_k=\overline{|x_k|}^{\,\alpha}$, so that they use more of
the grid, and divides the activations by $s_k$. Column scaling instead divides every weight column
by its own magnitude, $\widetilde W_{n,k}=W_{n,k}/c_k$, so that no outlier column sets the step
$(\max_{k\in g}\widetilde W_{n,k}-\min_{k\in g}\widetilde W_{n,k})/(2^b-1)$ of a 128-channel group;
the step then follows the typical columns, which lowers the rounding error of every weight in the
group, and the factor returns through the activation, $a'_k=a_kc_k$.

\subsection[Compensation-Aware Assignment]{Compensation-Aware Assignment}
\label{sec:algo:alloc}

\begin{algorithm}[t]
\caption{Proposed per-layer quantization}
\label{alg:quant}
\begin{algorithmic}[1]
\Require weights $W$, calibration Hessian $H$,  target width $b^\star$, share $p$
\State $c_k \gets \tfrac{1}{N}\sum_n |W_{n,k}|$;\quad
       $\widetilde W \gets W/c$;\quad fold $c$ into $H$
\State permute columns by descending $\mathrm{diag}(H)$ \Comment{act-order}
\State 
       $\rho_{n,g}\gets$ uncompensated error of a uniform $b^\star$-bit GPTQ pass (Eq.~\ref{eq:residual})
\State 
       $b^\star{+}1$ bits to the top-$p$ channels by $\rho$, $b^\star{-}1$ to the bottom-$p$, $b^\star$ to the rest
\For{each group $g$ in processing order}
  \State fit $(\text{scale}_{n,g},\text{zp}_{n,g})$  at its precision (W1: $\mu\pm\alpha$)
  \For{each column $k\in g$}
    \State $\hat w \gets$ quantize$(w_k)$ at the row's precision
    \State propagate $(w_k-\hat w)$ to remaining columns via $H^{-1}$
  \EndFor
\EndFor
\State \Return $q$, per-group $(\text{scale},\text{zp},\text{2-bit tag})$, $c$
\end{algorithmic}
\end{algorithm}
Our quantizer decides which output channels receive more bits by the error that
GPTQ \emph{cannot} compensate, rather than by a salience score computed before quantization, as in
SliM-LLM's salience-determined bit allocation (SBA)~\cite{slimllm} and PB-LLM~\cite{pbllm}. A
pre-quantization score misses two effects. GPTQ's error feedback already moves the rounding error of
salient columns onto less salient ones, an implicit bit allocation within each row, and column
scaling flattens the spread of group sensitivity ($4.97\times$ to $1.66\times$). A salience ranking
therefore promotes groups that GPTQ would have compensated anyway: SBA is no better than uniform 3-bit
(7.237 vs.\ 7.227 on Llama-3.1-8B, \autoref{tab:ablation}) when column scaling is enabled. We instead rank channels by the error that remains \emph{after} compensation. GPTQ's loss is a
sum of per-column terms $\big((w_{n,j}-q_{n,j})/U_{jj}\big)^2$, the part of each rounding error that
the remaining columns cannot absorb, where $U$ is the upper Cholesky factor of
$H^{-1}$~\cite{gptq}. A first GPTQ pass at the target width $b^\star$ gives each channel's share of this
loss in group $g$,
\begin{equation}
\label{eq:residual}
\rho_{n,g}=\sum_{j\in g}\Big(\frac{w_{n,j}-q^{(b^\star)}_{n,j}}{U_{jj}}\Big)^{2}.
\end{equation}
One more bit shrinks a channel's error, and with it $\rho_{n,g}$, by a roughly constant factor $r$, so
promoting a channel saves $(1-1/r)\rho_{n,g}$ and demoting one costs $(r-1)\rho_{n,g}$. For a given number
of channels moved, promoting those with the largest $\rho$ by one bit and demoting as many with the
smallest $\rho$ by one bit therefore minimizes the loss to first order while keeping the average at
exactly $b^\star$. A second GPTQ pass then quantizes the original weights at these widths
(Algorithm~\ref{alg:quant}). The number of channels moved is one global share ($1\%$ for F3, $20\%$ for
F2; \autoref{sec:setup}), so unlike SBA and AMQ~\cite{amq}, which search an allocation per layer, the
assignment needs no per-layer search; its cost is one extra GPTQ pass per layer.

\subsection{Storage Overhead}
\label{sec:algo:overhead}

Each layer stores the codes $q$ at their assigned widths, one FP16 scale and integer zero point
per group as in GPTQ, a $2$-bit precision tag per group, i.e., $0.016$
bits per weight, and the vector $c$: one FP16 value per input channel (below $0.004$ bits per
weight for $N\ge4096$) and one power-of-two exponent. The reported 3.0- and 2.0-bit averages
count the codes only, as is standard.

\section{Evaluation}

\subsection{Experimental Setup}
\label{sec:setup}

\begin{table}[b]
\centering\footnotesize
\setlength{\tabcolsep}{3.5pt}
\caption[Precision fractions]{Precision fractions at each operating point (same for all models; $p$
selected on the WikiText-2 validation split). MAC/cycle: effective rate of the mix on a 16-lane PE
(INT4/3/2/1: 8/10.67/16/32); $\times$INT4: speedup over uniform INT4; Util.: array utilization
of the sorted tile schedule (Section~\ref{sec:arch:accel}) on Llama-3.1-8B.}
\label{tab:fractions}
\resizebox{\columnwidth}{!}{%
\begin{tabular}{lcccccc}
\toprule
Point & $p$ & W4/W3/W2/W1 (\%) & Bits & MAC/cycle & $\times$INT4 & Util. \\
\midrule
F3 & $1\%$  & 1/98/1/--   & 3.00 & 10.67 & $1.33\times$ & 99.4\% \\
F2 & $20\%$ & --/20/60/20 & 2.00 & 16.0  & $2.00\times$ & 99.0\% \\
\bottomrule
\end{tabular}}
\end{table}

\betterparagraph{Quantization baselines and model quality}

\begin{table}[t]
\centering
\caption[Datapaths under comparison]{Datapaths under comparison; only \accelerator{} is accurate, bit-parallel, and precision-scaling.}
\label{tab:baselines}
\footnotesize
\begin{tabular}{ll}
\toprule
Design & Key datapath \\
\midrule
FP FMA               & FP mult.\ $+$ align $+$ normalize $\to$ FP acc. \\
Dot-product          & FP$\times$INT4, per-element normalize $\to$ FP acc. \\
FIGNA~\cite{figna}   & shared-exp.\ align $+$ INT mult.\ $\to$ INT acc. \\
FIGLUT~\cite{figlut} & activation LUT read (bit-serial) $\to$ INT acc. \\
AxCore~\cite{axcore} & log-approx.\ mult.\ $\to$ partial-FP acc. \\
\midrule
\textbf{\accelerator} & {radix-4 $4{:}1$ mux, no mult.\ $\to$ INT acc.} \\
\bottomrule
\end{tabular}
\end{table}

We evaluate six open models spanning three families and two size classes: LLaMA-2-7B~\cite{llama2},
Llama-3.1-8B~\cite{llama3}, Qwen2.5-7B~\cite{qwen2.5}, Qwen3-8B and Qwen3-32B~\cite{qwen3}, and
Gemma4-31B~\cite{gemma4}. We report WikiText-2~\cite{wikitext} and C4~\cite{c4} perplexity (2048-token
sequences; C4 on the GPTQ evaluation split) and zero-shot accuracy on PIQA~\cite{piqa},
ARC-easy/challenge~\cite{arc}, HellaSwag~\cite{hellaswag}, and Winogrande~\cite{winogrande}
(lm-eval-harness~\cite{lmeval}). Two operating points are used throughout, \textbf{F3} and
\textbf{F2} (3.00 and 2.00 average bits, \autoref{tab:fractions}); their share $p$
(\autoref{sec:algo:alloc}) is the best of a small grid on the WikiText-2 \emph{validation} split of
LLaMA-2-7B and is fixed for all models. We compare against GPTQ~\cite{gptq}, AWQ~\cite{awq},
SliM-LLM~\cite{slimllm}, AMQ~\cite{amq} (at its paper's search budget), and QuIP\#~\cite{quipsharp},
each with its official implementation. GPTQ, SliM-LLM, and \accelerator{} calibrate on the same 128
random 2048-token WikiText-2 training segments; AWQ, AMQ, and QuIP\# use their authors' calibration
sets (Pile~\cite{pile}, RedPajama~\cite{redpajama}). Only the decoder linear layers are quantized;
embeddings, the LM head, and activations stay in FP16, and no method is fine-tuned.

\begin{table*}[t]
\centering\footnotesize
\setlength{\tabcolsep}{1.2pt}
\caption[Perplexity and zero-shot accuracy]{WikiText-2 (Wiki2) and C4 perplexity ($\downarrow$) and
zero-shot accuracy (\%, $\uparrow$; HellaSwag (HS) length-normalized) at 3- and 2-bit average weight
precision. L2-7B: Llama-2-7B~\cite{llama2}; L3.1-8B: Llama-3.1-8B~\cite{llama3}; Q2.5-7B:
Qwen2.5-7B~\cite{qwen2.5}; Q3-8B/32B: Qwen3-8B/32B~\cite{qwen3}; G4-31B: Gemma4-31B~\cite{gemma4}.
$^{\ddagger}$AMQ runs at 3.03--3.05 and 2.25--2.40 average bits. AWQ diverges at 2 bits
(PPL $>\!10^{3}$) and is omitted there.}
\label{tab:ppl}
\resizebox{\textwidth}{!}{%
\begin{tabular}{l c
  w{c}{2.6em} *{5}{w{c}{2.6em}} w{c}{2.6em} @{\hspace{6pt}}
  w{c}{2.6em} *{5}{w{c}{2.6em}} w{c}{2.6em} @{\hspace{6pt}}
  w{c}{2.6em} *{5}{w{c}{2.6em}} w{c}{2.6em}}
\toprule
 & & \multicolumn{7}{c}{L2-7B} & \multicolumn{7}{c}{L3.1-8B} & \multicolumn{7}{c}{Q2.5-7B} \\
\cmidrule(lr){3-9}\cmidrule(lr){10-16}\cmidrule(lr){17-23}
Method & Bits & Wiki2 & C4 & PIQA & ARCe & ARCc & HS & Wino & Wiki2 & C4 & PIQA & ARCe & ARCc & HS & Wino & Wiki2 & C4 & PIQA & ARCe & ARCc & HS & Wino \\
\midrule
FP16 & 16 & 5.47 & 6.97 & 78.1 & 75.5 & 43.0 & 76.2 & 69.2 & 6.24 & 8.95 & 79.1 & 82.2 & 52.1 & 79.3 & 74.5 & 6.85 & 10.44 & 78.8 & 80.4 & 47.8 & 60.0 & 72.9 \\
\midrule
GPTQ & 3 & 6.14 & 8.05 & 76.0 & 73.1 & 39.0 & 71.9 & 67.6 & 7.74 & 11.95 & 75.5 & 72.8 & 40.2 & 73.9 & 70.7 & 7.69 & 11.77 & 76.9 & 77.1 & 45.9 & 55.8 & 68.1 \\
AWQ & 3 & 6.18 & 7.81 & 76.6 & 74.3 & 40.6 & 54.8 & 68.0 & 8.15 & 11.56 & 76.6 & 76.1 & 45.1 & 56.1 & 70.9 & 8.14 & 11.75 & 77.5 & 79.4 & 46.8 & 56.0 & 69.5 \\
QuIP\# & 3 & 6.24 & 7.70 & 76.5 & 74.4 & 41.2 & 53.4 & 69.1 & 7.79 & 10.89 & 77.3 & 75.3 & 43.6 & 56.1 & 70.8 & 7.45 & 11.05 & 78.4 & 80.4 & 48.6 & 56.9 & 72.0 \\
AMQ$^{\ddagger}$ & 3.05 & 6.70 & 8.83 & 76.4 & 69.4 & 42.4 & 72.1 & 68.7 & 9.20 & 14.29 & 76.6 & 70.8 & 43.9 & 71.3 & 70.0 & 12.30 & 20.41 & 73.9 & 66.5 & 41.5 & 67.0 & 63.3 \\
SliM-LLM & 3 & 6.40 & 8.57 & 75.4 & 52.2 & 37.4 & 69.1 & 64.5 & 7.70 & 13.71 & 77.2 & 68.4 & 44.5 & 73.9 & 71.1 & 7.60 & 13.43 & 77.5 & 72.3 & 48.3 & 75.4 & 69.1 \\
\textbf{\accelerator} & 3 & \textbf{5.85} & \textbf{7.67} & \textbf{76.3} & \textbf{74.5} & \textbf{42.3} & \textbf{73.3} & \textbf{68.7} & \textbf{7.19} & \textbf{10.95} & \textbf{78.1} & \textbf{78.2} & \textbf{46.8} & \textbf{76.0} & \textbf{71.3} & \textbf{7.46} & \textbf{11.37} & \textbf{78.8} & \textbf{79.1} & \textbf{46.1} & \textbf{57.0} & \textbf{70.6} \\
\midrule
GPTQ & 2 & 32.61 & 59.80 & 53.2 & 25.6 & 21.5 & 25.6 & 48.1 & 98.94 & 310.8 & 54.2 & 26.1 & 20.5 & 27.5 & 50.9 & 19.62 & 60.83 & 56.8 & 33.9 & 24.8 & 31.4 & 50.4 \\
QuIP\# & 2 & 13.24 & 12.83 & 71.1 & 61.7 & 30.0 & 41.2 & 62.8 & 21.20 & 23.35 & 69.1 & 65.1 & 21.2 & 39.7 & 62.7 & 14.16 & 15.25 & 74.1 & 76.6 & 42.1 & 47.6 & 67.6 \\
AMQ$^{\ddagger}$ & 2.40 & 10.60 & 13.99 & 72.1 & 55.1 & 35.2 & 62.4 & 61.6 & 22.90 & 33.87 & 70.2 & 54.3 & 32.4 & 54.2 & 62.3 & 44.40 & 62.04 & 66.7 & 55.0 & 32.9 & 54.5 & 58.4 \\
SliM-LLM & 2 & 30.80 & 36.83 & 62.0 & 39.1 & 27.1 & 39.8 & 55.6 & 63.10 & 298.76 & 54.8 & 32.5 & 31.3 & 29.4 & 50.8 & 15.00 & 48.54 & 61.8 & 47.0 & 28.2 & 46.1 & 53.9 \\
\textbf{\accelerator} & 2 & \textbf{8.77} & \textbf{13.86} & \textbf{66.9} & \textbf{54.5} & \textbf{28.7} & \textbf{55.2} & \textbf{59.3} & \textbf{13.77} & \textbf{27.36} & \textbf{64.2} & \textbf{50.6} & \textbf{26.1} & \textbf{49.4} & \textbf{59.6} & \textbf{10.47} & \textbf{17.43} & \textbf{67.7} & \textbf{57.9} & \textbf{30.1} & \textbf{45.1} & \textbf{60.1} \\
\midrule\midrule
 & & \multicolumn{7}{c}{Q3-8B} & \multicolumn{7}{c}{Q3-32B} & \multicolumn{7}{c}{G4-31B} \\
\cmidrule(lr){3-9}\cmidrule(lr){10-16}\cmidrule(lr){17-23}
Method & Bits & Wiki2 & C4 & PIQA & ARCe & ARCc & HS & Wino & Wiki2 & C4 & PIQA & ARCe & ARCc & HS & Wino & Wiki2 & C4 & PIQA & ARCe & ARCc & HS & Wino \\
\midrule
FP16 & 16 & 9.72 & 13.29 & 76.9 & 83.6 & 55.8 & 57.1 & 67.8 & 7.60 & 10.77 & 81.1 & 84.3 & 57.9 & 63.9 & 73.1 & 6.38 & 9.29 & 82.9 & 88.4 & 64.9 & 65.9 & 80.0 \\
\midrule
GPTQ & 3 & 10.66 & 14.64 & 74.3 & 76.7 & 47.5 & 53.2 & 65.8 & 8.48 & 11.88 & 77.7 & 78.5 & 51.5 & 61.2 & 67.9 & 7.23 & 10.72 & 81.6 & 83.7 & 56.1 & 61.8 & 76.9 \\
AWQ & 3 & 11.48 & 15.08 & 75.1 & 78.2 & 47.8 & 52.8 & 65.0 & 8.63 & 11.74 & 79.3 & 81.1 & 53.7 & 61.2 & 72.5 & 8.26 & 11.43 & 81.3 & 86.2 & 61.3 & 62.6 & 77.7 \\
QuIP\# & 3 & 10.41 & 14.20 & 75.8 & 81.9 & 52.2 & 53.9 & 68.4 & 8.04 & 11.18 & 80.0 & 83.0 & 59.3 & 62.3 & 72.1 & 7.21 & 10.13 & 82.2 & 86.6 & 60.6 & 62.4 & 79.0 \\
AMQ$^{\ddagger}$ & 3.05 & 12.90 & 19.35 & 74.8 & 74.6 & 49.7 & 68.2 & 64.0 & 9.13 & 14.24 & 80.2 & 79.7 & 57.7 & 79.2 & 68.4 & 8.61 & 13.28 & 82.6 & 83.8 & 60.0 & 80.8 & 77.3 \\
SliM-LLM & 3 & 11.60 & 18.54 & 75.5 & 70.4 & 48.4 & 69.9 & 65.2 & 8.41 & 13.78 & 80.4 & 78.5 & 58.2 & 80.2 & 71.4 & 7.20 & 11.57 & 83.4 & 86.1 & 62.5 & 82.1 & 77.0 \\
\textbf{\accelerator} & 3 & \textbf{10.30} & \textbf{14.24} & \textbf{74.7} & \textbf{80.1} & \textbf{50.1} & \textbf{53.5} & \textbf{66.8} & \textbf{8.11} & \textbf{11.48} & \textbf{78.5} & \textbf{81.7} & \textbf{54.1} & \textbf{61.5} & \textbf{71.4} & \textbf{6.99} & \textbf{10.27} & \textbf{81.9} & \textbf{85.9} & \textbf{59.1} & \textbf{62.8} & \textbf{77.5} \\
\midrule
GPTQ & 2 & 35.21 & 74.14 & 51.9 & 27.8 & 22.4 & 27.9 & 51.3 & 23.31 & 34.90 & 57.9 & 31.8 & 22.8 & 36.3 & 49.3 & 53.97 & 150.67 & 58.4 & 36.8 & 23.9 & 33.7 & 51.9 \\
QuIP\# & 2 & 21.18 & 21.02 & 71.1 & 75.3 & 43.0 & 45.4 & 65.5 & 9.59 & 13.05 & 76.3 & 83.5 & 55.8 & 57.1 & 72.5 & 13.01 & 15.36 & 76.7 & 80.9 & 48.7 & 49.5 & 74.1 \\
AMQ$^{\ddagger}$ & 2.40 & 32.30 & 41.65 & 69.0 & 51.9 & 32.3 & 53.0 & 57.6 & 11.94 & 18.00 & 76.1 & 74.5 & 48.9 & 70.6 & 64.3 & 36.19 & 47.89 & 74.3 & 64.6 & 40.0 & 64.5 & 62.0 \\
SliM-LLM & 2 & 26.50 & 57.24 & 62.0 & 39.2 & 26.5 & 43.1 & 50.9 & 13.90 & 58.11 & 66.8 & 46.5 & 32.5 & 61.9 & 58.0 & 14.10 & 28.85 & 68.2 & 51.5 & 33.4 & 53.0 & 61.3 \\
\textbf{\accelerator} & 2 & \textbf{14.62} & \textbf{23.96} & \textbf{65.8} & \textbf{52.2} & \textbf{27.7} & \textbf{39.4} & \textbf{56.8} & \textbf{11.20} & \textbf{17.02} & \textbf{72.0} & \textbf{60.4} & \textbf{34.6} & \textbf{50.1} & \textbf{61.5} & \textbf{10.95} & \textbf{17.99} & \textbf{75.5} & \textbf{71.7} & \textbf{39.8} & \textbf{47.8} & \textbf{66.1} \\
\bottomrule
\end{tabular}}
\end{table*}

\betterparagraph{Hardware Baselines}
We implement floating-point Fused Multiply-Add (FMA) and FP dot-product
engines as conventional baseline engines. Beyond them, we compare \accelerator{} against three state-of-the-art FP-INT accelerators, all
re-implemented and synthesized in our own flow. \textbf{FIGNA}~\cite{figna}
aligns each activation block to a shared exponent and multiply-accumulates the
aligned significands against the INT4 weights as integers, returning to floating
point once per output. \textbf{AxCore}~\cite{axcore} is approximate: a shared
pre-add applies the Mitchell compensation bias, the FP$\times$INT multiply
reduces to a logarithmic-domain integer addition, and the result is accumulated in a per-PE partial floating-point adder that defers
normalization and rounding to a shared module, as in the published design. \textbf{FIGLUT}~\cite{figlut} replaces the
multiplier with a table read, pre-computing sums of aligned activations once and
reading one entry per weight digit; we implement the half-size LUT of its paper
as real register storage and take its scaling with activation width from
the published cross-format comparison. \autoref{tab:baselines} summarizes and compares the key datapaths of each baseline. Our re-implementations track the published area ratios: averaged over FP16 and FP32, our
AxCore-to-FIGNA ratio is within 2.5 points of the AxCore paper and our FIGLUT-to-FIGNA ratio within
$2.4\%$ of the FIGLUT paper.

\betterparagraph{Hardware evaluation setup}
We implement \accelerator{} and all baselines as parameterized SystemVerilog RTL and
synthesize them using Synopsys Design Compiler in the NanGate
$15$\,nm technology, targeting $1$\,GHz. All results use this common clock. Every design meets it in all three
activation formats, and all but the dot-product engine (1\,ps of slack in FP32) close with margin, so
the common clock does not favor \accelerator{}. For a fair comparison,
all designs are adjusted to the same maximum-performance (TOPS) target on a
$128\!\times\!128$ array ($16{,}384$ MAC/cycle). All designs are evaluated for nine
cases---activations in FP32, FP16, or BF16 and weights in INT4, INT3, or INT2---with each
design fully re-synthesized per activation format rather than analytically scaled. INT4 is
a uniform 4-bit quantization; INT3 and INT2 are our proposed mixed-precision quantization averaging $3.0$
and $2.0$ bits per weight (the F3 and F2 mixes of \autoref{tab:fractions}). Rigid baselines process any sub-INT4 weight by
zero-padding their INT4 datapath, as published, and so gain nothing below 4 bits. Power is obtained under an identical switching activity of $0.2$ toggles per clock
cycle for every design.
To model memory properly,
input, weight, and output buffers are compiled as SRAM and characterized with
CACTI~\cite{cacti}, sized identically across designs. They hold $1.2$\,MB in total. Off-chip memory
is one HBM3E stack at $3.44$\,pJ/bit~\cite{hbm3e_energy} with
$1280$\,GB/s~\cite{hbm3e_isscc}; its traffic follows from tiling each GEMM with a
$512\times512$ on-chip output block and is the same per MAC for every design, so HBM power differs
only through throughput. Total energy comprises three terms: (1) memory energy, the access count times the
energy per bit, (2) logic dynamic energy, the synthesized per-operation energy times the
operation count, and (3) leakage energy, the leakage power times the latency. Following FIGNA~\cite{figna}, we report TOPS/mm$^2$ and normalized energy, including SRAM and off-chip communication. Results are given for four open LLMs (Llama-3-8B, Qwen3-8B, Qwen3-32B, Gemma4-31B), each normalized to the same-format, same-precision floating-point FMA engine.

\begin{figure}[t]
\centering
\includegraphics[width=\linewidth]{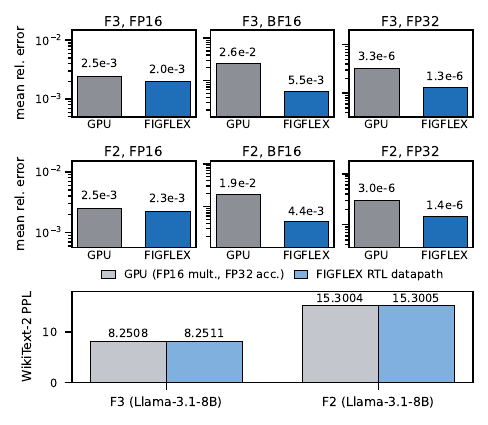}
\vspace{-7mm}
\caption{\small Numerical fidelity on the F3 and F2 models (Llama-3.1-8B).
\emph{Top:} mean relative error against the float64 dot product over 21 projections; the tensor core
multiplies in the activation format and accumulates in FP32. \emph{Bottom:} WikiText-2 perplexity on
the GPU and through the \accelerator{} datapath.}
\vspace{-3mm}
\label{fig:numerics_f3f2}
\end{figure}

\subsection{Numerical Fidelity}
\label{sec:fidelity}

To verify that model quality is preserved on real end-to-end requests, we run the quantized F3
and F2 Llama-3.1-8B models through a bit-accurate model of the \accelerator{} datapath, including the
column scales, the shared-exponent alignment (6 guard bits, truncation, subnormals flushed), the
integer accumulation, and the group epilogue. Their WikiText-2 perplexity matches the GPU to the third
decimal (F3: 8.2511 vs.\ 8.2508; F2: 15.3005 vs.\ 15.3004 on four test windows), so the quality in
Table~\ref{tab:ppl} holds on \accelerator{}. We validate the model against the RTL: the integer
datapath of the full $16\times128$-PE array, simulated in Verilator on all 21 projections of blocks
0, 15, and 31 with real activations and each group in its own mode, matches it bit for bit on all
$10.0$ million group results; the FP32 post-processing uses DesignWare units modeled from their
specification.

\autoref{fig:numerics_f3f2} compares the per-projection error against the float64 dot product.
The radix-4 recoding and the integer accumulation are lossless, so the only rounding is the
shared-exponent alignment, and \accelerator{} is at least as accurate as a tensor core that
multiplies in the activation format and accumulates in FP32, for FP16, BF16, and FP32 activations
(FP16, F3: $2.0\times10^{-3}$ vs.\ $2.5\times10^{-3}$). Finer exponent blocks or 24 guard bits leave
the error unchanged, and the power-of-two rescaling (\autoref{sec:algo:colscale}) keeps the
column-scaled activations out of the FP16 subnormal range.

\subsection{Model Quality}
\label{sec:quality}

Table~\ref{tab:ppl} reports perplexity and zero-shot accuracy at 3- and 2-bit average precision.
GPTQ~\cite{gptq}, AWQ~\cite{awq}, SliM-LLM~\cite{slimllm}, AMQ~\cite{amq}, and \accelerator{} produce uniform affine integer codes that an FP-INT
array consumes directly. 
QuIP\#~\cite{quipsharp} is run without fine-tuning and is a reference for quality without the affine-grid constraint, not a ceiling.

\noindent\textbf{Perplexity.}
At 3 bits, \accelerator{} has the lowest WikiText-2 perplexity among the affine methods on all six
models and the lowest overall on four; QuIP\# leads on the other two by at most $0.07$. At 2 bits,
where a uniform affine grid normally fails, the gap widens: GPTQ, AWQ, and SliM-LLM collapse, while
\accelerator{} stays within $1.5$--$2.2\times$ of FP16 and is the best method overall on five of six
models, $21\%$ below QuIP\# on average and below AMQ on all six despite AMQ's $2.40$-bit average. On
the held-out C4, \accelerator{} remains the best affine method on every model at both widths and is
within $0.32$ of QuIP\# at 3 bits.

\noindent\textbf{Zero-shot accuracy.} Averaged over PIQA, ARC-e, ARC-c, and Winogrande on the six
models (FP16: 72.1\%), \accelerator{} is the most accurate affine method at 3 bits (69.7\%,
vs.\ 69.2\% for AWQ and 67.5\% for GPTQ), within 0.7 points of QuIP\#. At 2 bits it leads
GPTQ and SliM-LLM at the same 2.00 bits by 14.9 and 7.0 points, and trails AMQ by 2.6
points and QuIP\# by 9.1.

\subsection{Quantizer Ablation}
\label{sec:ablation}

\autoref{tab:ablation} adds the components of our quantizer one at a time on Llama-3.1-8B with the
same calibration. The tuned uniform row adds clip search and column scaling to GPTQ; the next rows
add our compensation-aware (CA) mixed assignment without and with column scaling, and SBA-Mixed
replaces it with SliM-LLM's salience-determined bit allocation~\cite{slimllm} (a pre-quantization
ranking $w^2/U_{jj}^2$ with a per-layer search of the share) on the same backbone. Latency is the linear-GEMM time of one
2048-token prefill on the iso-TOPS array, including the column-scale multiply and the measured tile
utilization (\autoref{tab:fractions}).

At 3 bits, CA-Mixed lowers perplexity at $0.6\%$ higher latency. The gain is small because
3-bit error is already low; F3's hardware benefit is throughput instead (10.67 vs.\ 8 MAC/cycle for
INT4). At 2 bits, mixed precision is the larger effect: on its own, CA-Mixed lowers perplexity from
98.94 (GPTQ) to 16.48, below the tuned uniform model (29.28), and column scaling brings it to 13.77.
The ranking matters: SBA-Mixed is no better than uniform at 3 bits (7.237 vs.\ 7.227) and reaches
only 18.84 at 2 bits, because a salience score computed before quantization cannot see which groups
GPTQ will compensate. Because every precision runs at $32/b$ MAC/cycle, equal average bits
mean equal throughput: F2 runs at the rate of uniform W2 up to its $1\%$ tile-packing loss (453.2
vs.\ 448.8\,ms) at $2.1\times$ lower perplexity. Column scaling costs 12.6\,ms per prefill
(1.9--2.8\%).

\begin{table}[t]
\centering\footnotesize
\setlength{\tabcolsep}{4pt}
\caption[Quantizer ablation]{Quantizer ablation on Llama-3.1-8B (FP16: 6.238 WikiText-2 PPL). Every
row after GPTQ builds on it (act-order GPTQ). CA: compensation-aware; SBA: salience-determined bit
allocation. Latency: linear-GEMM time of one 2048-token prefill, including the column-scale multiply.}
\label{tab:ablation}
\resizebox{\columnwidth}{!}{%
\begin{tabular}{lcccc}
\toprule
 & \multicolumn{2}{c}{WikiText-2 PPL} & \multicolumn{2}{c}{GEMM latency (ms)} \\
\cmidrule(lr){2-3}\cmidrule(lr){4-5}
Configuration & 3-bit & 2-bit & 3-bit & 2-bit \\
\midrule
GPTQ (official, act-order)               & 7.744 & 98.94 & 654.3 & 436.2 \\
Tuned uniform (col.\ scaling) & 7.227 & 29.28 & 666.9 & 448.8 \\
CA-Mixed, no col.\ scaling & 7.370 & 16.48 & 658.3 & 440.6 \\
SBA-Mixed + col.\ scaling~\cite{slimllm} & 7.237 & 18.84 & 670.8 & 453.2 \\
\textbf{CA-Mixed + col.\ scaling (ours)} & \textbf{7.193} & \textbf{13.77} & 670.8 & 453.2 \\
\bottomrule
\end{tabular}}
\end{table}

\subsection{Hardware Efficiency}
\label{sec:hw}
All hardware results in this section are for the linear GEMM layers, which dominate LLM
runtime (\autoref{fig:motivation}); attention and the rest of the model are outside FIGFLEX. We
estimate the end-to-end effect at the end of the section.

\begin{table}[t]
\centering\footnotesize
\setlength{\tabcolsep}{3pt}
\caption[Datapath ablation]{Datapath ablation, synthesized in the same flow (FP16 activations).
Op.: precomputed operands per lane; PE area includes registers; Repl.: row-replicator area per PE;
area and power per MAC at the INT4 rate. All designs run INT4/INT2/INT1 at 8/16/32 MAC/cycle.}
\label{tab:radix}
\setlength{\tabcolsep}{2.5pt}
\begin{tabular}{lcccccc}
\toprule
Datapath & Op. & PE & Repl. & $\mu$m$^2$/ & mW/ & INT3 \\
 & & ($\mu$m$^2$) & ($\mu$m$^2$) & MAC & MAC & MAC/cyc. \\
\midrule
Radix-2 tree        & 1  & 1144 & 12.9 & 144.6 & 0.167 & 10.67 \\
Radix-4             & 3  & 873  & 12.7 & 110.8 & 0.162 & 8 \\
\textbf{Radix-4+INT3} & 3 & 881 & 16.3 & \textbf{112.2} & \textbf{0.165} & 10.67 \\
Radix-8             & 7  & 911  & 21.3 & 116.6 & 0.168 & 10.67 \\
Radix-16            & 15 & 1239 & 30.7 & 158.7 & 0.201 & 10.67 \\
\bottomrule
\end{tabular}
\end{table}

\betterparagraph{Datapath ablation}
\autoref{tab:radix} isolates the datapath choices, each synthesized in the same flow. Moving
from a radix-2 fused adder tree to radix-4 with the multiples hoisted to the row edge cuts the PE
area by $24\%$ and the area per MAC by $23\%$ at the same power per MAC, because each lane resolves
two weight bits with one $4{:}1$ mux and the adder tree halves. The INT3 fusion costs $1.3\%$ area
(the replicator grows from 12.7 to 16.3\,$\mu$m$^2$ per PE) and raises the INT3 rate from 8 to 10.67
MAC/cycle, so the area per INT3 MAC drops from 110.8 to 84.1\,$\mu$m$^2$ ($1.32\times$). Higher radices
need more precomputed operands per lane (7 for radix-8, 15 for radix-16), whose replicator and wider
muxes outweigh the smaller tree: radix-8 costs $3.9\%$ more area and $1.8\%$ more power per MAC than
our datapath, and radix-16 $41\%$ and $22\%$. Radix-4 is therefore the sweet spot, and the INT3
fusion is nearly free. Because the INT4 comparison below involves no precision flexibility, it
reflects this datapath alone.

\begin{table}[t]
\centering\footnotesize
\setlength{\tabcolsep}{2.4pt}
\caption[Absolute hardware results]{Absolute results (FP16 activations, $128\times128$ MACs, 1\,GHz).
Area includes the same 1.2\,MB SRAM (1.28\,mm$^2$) and output conversion (0.52\,mm$^2$). Power at
INT4 (0.2 toggle rate), chip and with HBM3E traffic (Llama-3.1-8B). All designs reach 32.8\,TOPS and a
4.76\,s request at INT4. Request: one Llama-3.1-8B chat request (4k/4k tokens, batch 128; time per
request), weights and KV cache at the F3/F2 bit-widths, HBM3E at 1280\,GB/s, attention on a common
FP16 unit.}
\label{tab:abs}
\begin{tabular}{lccccccc}
\toprule
 & Area & \multicolumn{2}{c}{Power (W)} & \multicolumn{2}{c}{TOPS} & \multicolumn{2}{c}{Request (s)} \\
\cmidrule(lr){3-4}\cmidrule(lr){5-6}\cmidrule(lr){7-8}
Design & (mm$^2$) & chip & +HBM & F3 & F2 & F3 & F2 \\
\midrule
FP FMA      & 11.10 & 17.52 & 20.17 & 32.8 & 32.8 & 4.58 & 4.39 \\
Dot-product & 9.52  & 9.59  & 12.24 & 32.8 & 32.8 & 4.58 & 4.39 \\
FIGNA~\cite{figna}   & 5.11 & 5.04 & 7.69 & 32.8 & 32.8 & 4.58 & 4.39 \\
FIGLUT~\cite{figlut} & 4.68 & 5.22 & 7.87 & 43.7 & 65.5 & 3.70 & 2.65 \\
AxCore~\cite{axcore} & 3.97 & 5.02 & 7.67 & 32.8 & 32.8 & 4.58 & 4.39 \\
\midrule
\textbf{\accelerator} & \textbf{3.43} & \textbf{4.02} & \textbf{6.67} & 43.4 & 64.9 & \textbf{3.72} & \textbf{2.67} \\
\bottomrule
\end{tabular}
\end{table}

\begin{figure*}[h]
\centering
\includegraphics[width=\textwidth]{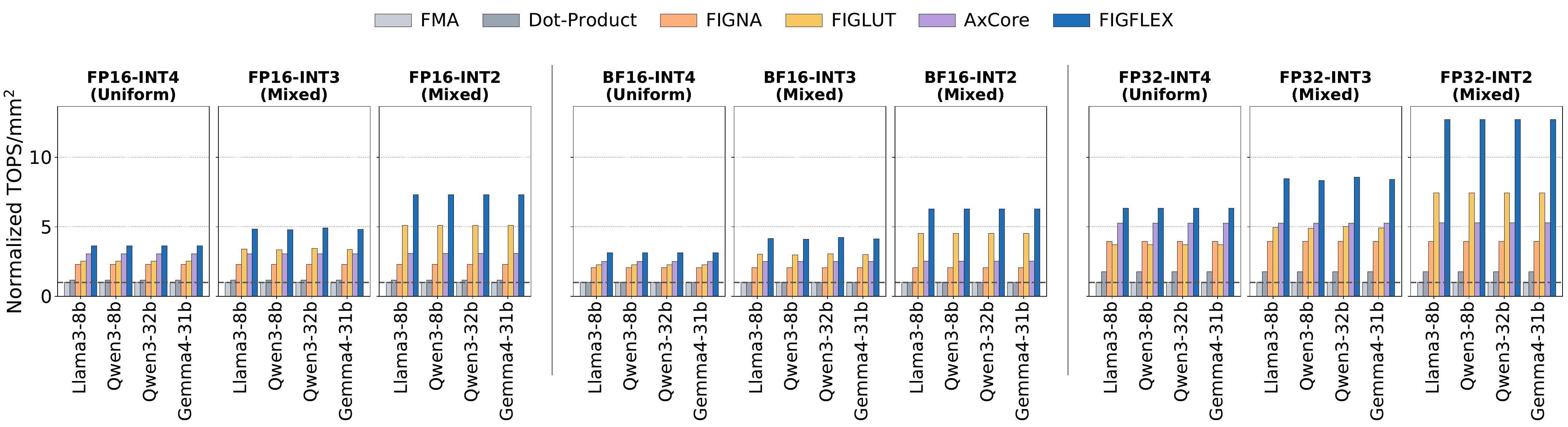}
\vspace{-7mm}
\caption{\small Compute density (TOPS/mm$^2$) normalized to a same-format FMA core
($\equiv 1.0$), for INT4 and mixed-precision (INT3) operating points across models.
\accelerator leads every design; its advantage grows at sub-4-bit precision.}
\vspace{-4mm}
\label{fig:tops_per_area}
\end{figure*}

\begin{figure*}[h]
\centering
\includegraphics[width=\textwidth]{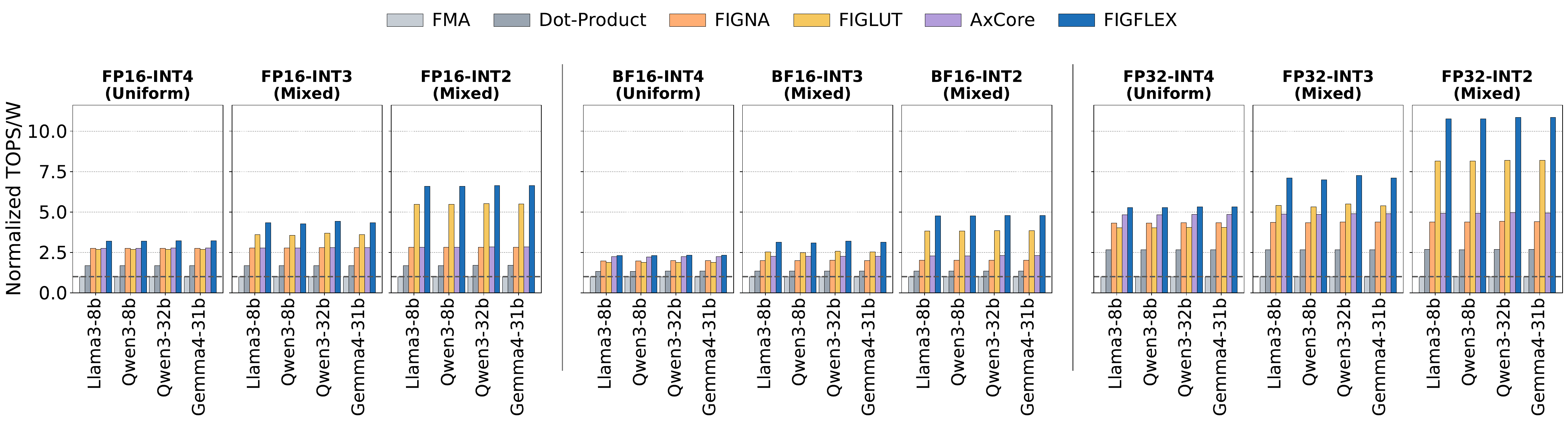}
\vspace{-7mm}
\caption{\small Energy efficiency (TOPS/W) normalized to the same-format FMA core.
\accelerator's integer, multiplier-free datapath is the most energy-efficient, ahead of
FIGNA's integer multiplier, FIGLUT's LUT reads, and AxCore's log-approximation.}
\vspace{-4mm}
\label{fig:tops_per_watt}
\end{figure*}

\betterparagraph{Absolute results}

\autoref{tab:abs} gives the absolute numbers behind the normalized figures. The physical units
differ: \accelerator{} has 2{,}048 PEs of 16 radix-4 lanes, FIGLUT 16{,}384 read-accumulate units
sharing about 1{,}107 8-entry LUTs, and the others 16{,}384 MAC units. \accelerator{} has the smallest
area and power because its PE has no multiplier and the activation multiples are formed once per
row; its compute array is $25\%$ smaller than AxCore's. At INT4 every design takes the same request
time, since the configuration is iso-TOPS. At F3 and F2, \accelerator{} is $1.23\times$ and
$1.64\times$ faster than the fixed-precision designs, which gain only from the smaller weights and KV
cache; FIGLUT, the other precision-scalable design, matches this runtime but needs $1.36\times$ the
area and $1.30\times$ the power. Power uses a common 0.2 toggle rate as an estimate. As a check, we replaced it with per-bit toggle rates and static
probabilities measured on real Llama-3.1-8B tensors in each design's input encoding, propagated by
the synthesis tool: at INT4, \accelerator's TOPS/W lead remains $1.18$--$1.50\times$ over FIGNA,
$1.15$--$1.46\times$ over AxCore, and $1.04$--$1.32\times$ over FIGLUT.

\betterparagraph{Compute density}

Figures~\ref{fig:tops_per_area} and~\ref{fig:tops_per_watt} report whole-chip compute density
(TOPS/mm$^2$) and energy efficiency (TOPS/W), including the SRAM, normalized to the same-format FMA design.
INT4 is uniform GPTQ; INT3 and INT2 use our mixed-precision quantizer. \accelerator{} is the most
area-efficient design at every operating point. At INT4, the operating point FIGNA and AxCore are
designed for and where no design benefits from precision flexibility, it reaches $1.49\times$ the
throughput per area of FIGNA and $1.18\times$ that of AxCore (geometric mean over the three formats),
which isolates the benefit of the radix-4 datapath. At INT3 the advantage grows to $1.97\times$ and
$1.56\times$, partly because those baselines are fixed-precision. Re-implementing them for 3 or 2 bits
would recover only part of this. In AxCore, narrower weights shrink only the approximate multiplier, a
small part of its area; the FP accumulator that follows dominates and stays intact
(\autoref{fig:area_comp}), so its throughput per area barely improves. Against a custom INT3/INT2
FIGNA, \accelerator's advantage would be similar to the uniform INT4 case, since it comes from the
fused radix-4 MAC datapath. Against designs that also scale with
precision, it remains: a BitBlade-style~\cite{bitblade} radix-2 tree (\autoref{tab:radix}) runs every
precision at \accelerator's rate but needs $1.29\times$ more area per MAC, and FIGLUT, which scales
throughput with precision as \accelerator{} does, trails it by $1.42\times$ at INT3 and INT2 alike,
purely from the per-MAC cost. Because every precision runs at $32/b$ MAC/cycle, throughput, density,
and energy per operation depend only on the average bit-width, not on how a quantizer distributes it.

\accelerator is also the most energy-efficient design at every datapath and operating point, surpassing even FIGLUT---which, being bit-serial, is inherently the most energy-efficient of the baselines below 4 bits---by an average of $25\%$.

\begin{figure}[t]
\centering
\includegraphics[width=\linewidth]{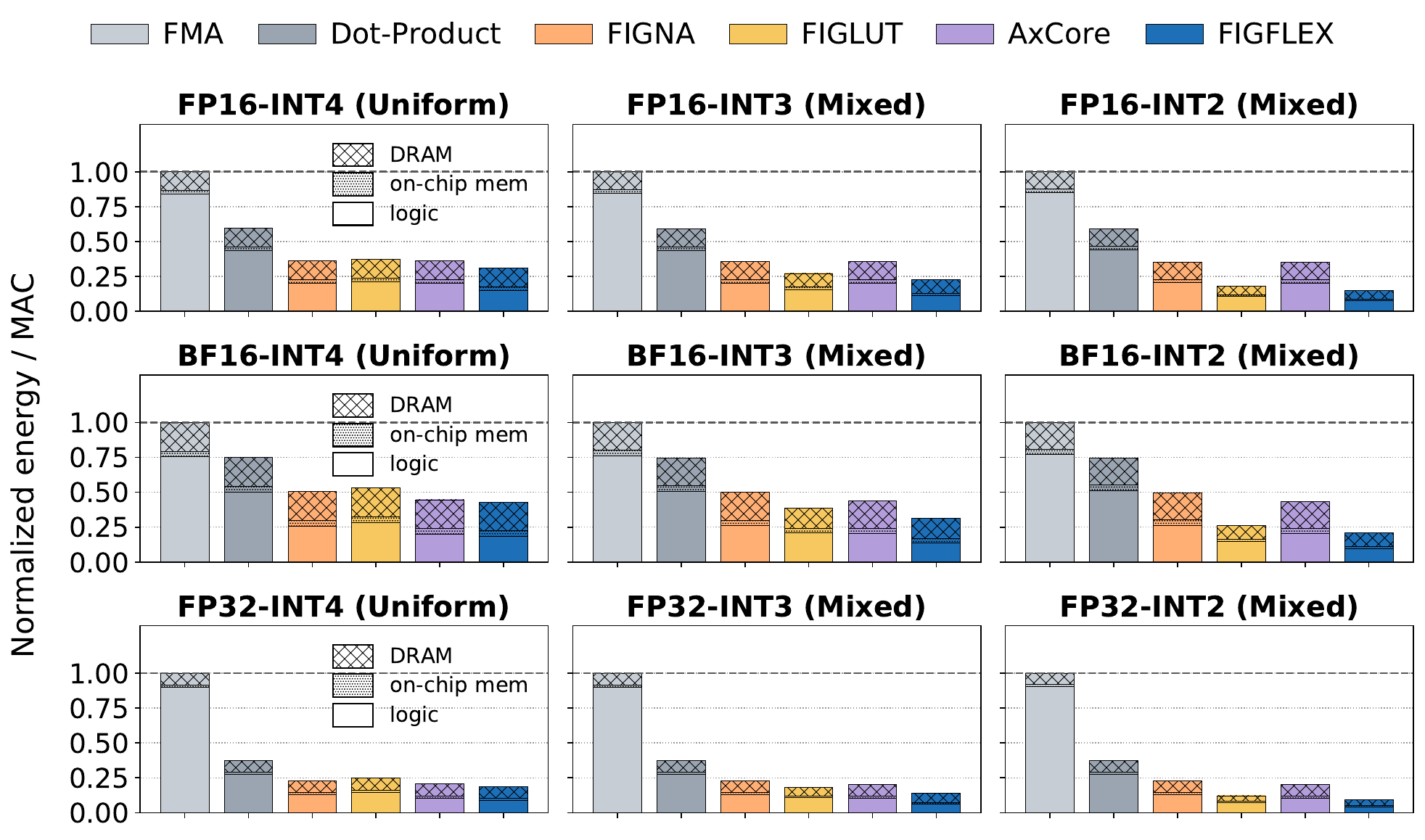}
\vspace{-7mm}
\caption{\small Normalized energy comparison of baseline accelerators running Qwen3-32B}
\vspace{-3mm}
\label{fig:energy_norm}
\end{figure}

\betterparagraph{Energy Comparison}
Figure~\ref{fig:energy_norm} compares the energy consumption of the baselines running three variations of quantization of the Qwen3-32B model (Uniform INT4, Mixed Precision INT3, and Mixed Precision INT2). \accelerator consumes the least energy at every operating point; AxCore is closest at INT4 and FIGLUT at INT3 and INT2.
DRAM and SRAM access energy is also captured, as discussed in Section~\ref{sec:setup}.

\betterparagraph{End-to-end estimate}

\autoref{fig:motivation} extends the evaluation to a full request with the first-order model of
Section~\ref{sec:bg:compute}. Only the linear GEMMs are measured for \accelerator{}: we place it in
the area of the GPU's tensor cores, where it runs the GEMMs at its measured compute density relative
to an FP16 dot-product engine ($3.2\times$ at INT4, \autoref{fig:tops_per_area}) times the F3/F2 rates
(\autoref{tab:fractions}), i.e., $4.2\times$ and $6.3\times$ the tensor-core rate. Attention, the
non-GEMM layers, and memory traffic are modeled identically for both systems: GPU-equivalent compute
and the same bandwidth, with weights and KV cache at the same bit-width. The baseline is the tensor
cores running low-bit kernels (W3A16 and W2A16, as in FLUTE~\cite{flute}), which dequantize to FP16
and compute at the FP16 rate. Replacing them with \accelerator{} shortens a batched request (batch
128) by $2.2$--$2.9\times$ at F3 and $2.9$--$3.7\times$ at F2. At batch 1, decode is memory-bound and
both read the same bytes, so the gain shrinks to $1.02$--$1.16\times$ and $1.04$--$1.27\times$.

\betterparagraph{Area and power breakdown}
Figures~\ref{fig:area_comp} and~\ref{fig:power_comp} break down accelerator
area and power excluding SRAM, across the compared engines; the SRAM, identical for every design (1.28\,mm$^2$), is included in \autoref{tab:abs}. The column-scale multiply of \autoref{sec:algo:colscale} reuses the per-channel scale multiply that the activation producer already performs (e.g., RMSNorm's $\gamma$), so it adds no area; its energy, one FP16 multiply (0.36\,pJ) per activation element shared by all $N$ output channels, is below $0.1\%$ of the per-MAC energy at $N{=}4096$. Because \accelerator{} hoists the
radix-4 generation logic to the shared per-row aligner and
replicator, its per-PE cost reduces to a select/AND bank, a binary adder tree, and an
integer accumulator, without adding arithmetic error. \accelerator{} is 25\% smaller than AxCore, the state-of-the-art 4-bit quantized accelerator. While it has a larger multiplication unit, our significantly smaller integer accumulator yields less area overall. Note that \accelerator{}'s multiplication area is obtained by adding the selection muxes and the proportion of the binary adder tree that performs partial-product reduction for INT4 quantization (lvl.\ 0 adders).
Comparing power consumption, \accelerator{} requires about 20--23\% less power than the other baselines. FIGLUT has no multiplier; the majority of the power is consumed by the mux, output-generation logic, and bit-wise accumulation. AxCore is also equipped with an efficient multiplier approximator, which is essentially a fixed integer adder, and the majority of the computation power is consumed by the FP accumulator. The power breakdown of \accelerator{} is obtained similarly to the area breakdown.

\begin{figure}[t]
\centering
\includegraphics[width=\linewidth]{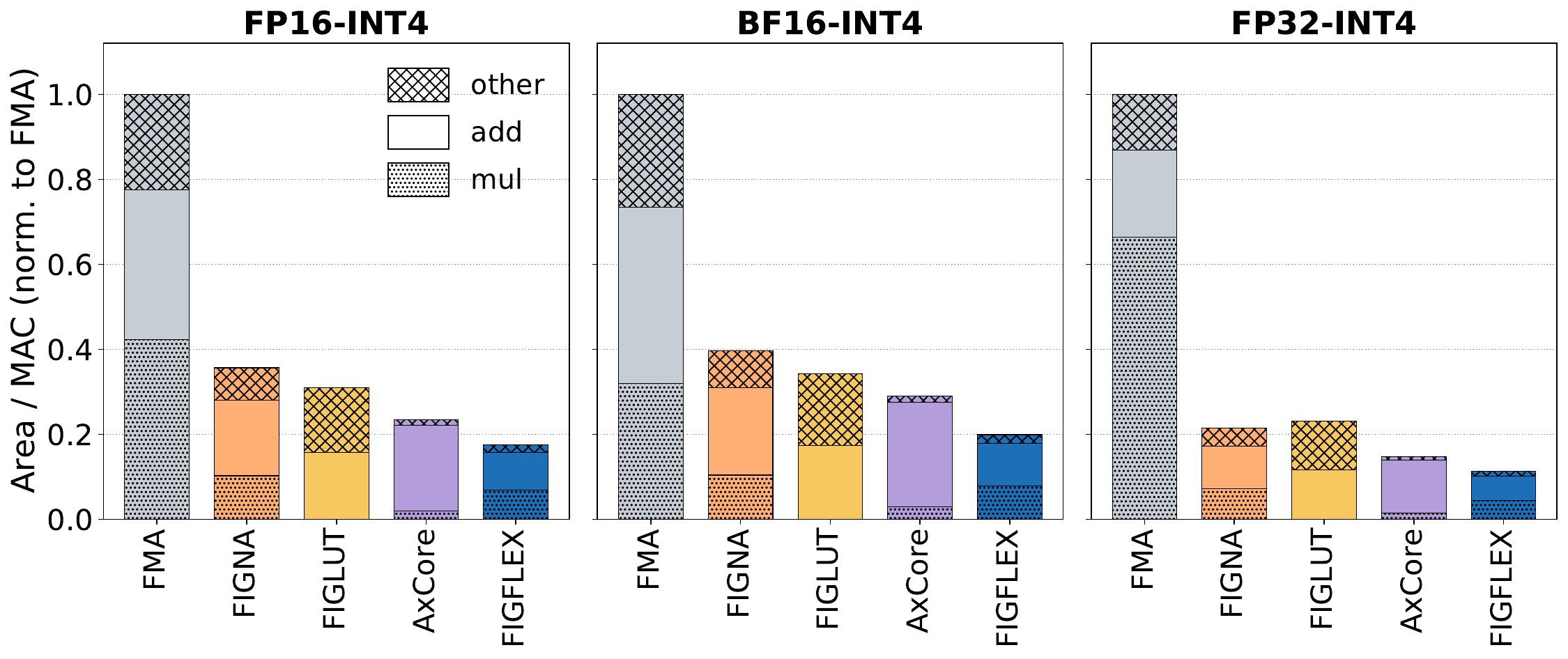}
\vspace{-7mm}
\caption{\small Accelerator area breakdown (excluding SRAM) across the compared engines.
\accelerator's alignment and multiple-generation shifters are amortized at the shared
array edge, leaving lean select/adder-tree PEs.}
\vspace{-3mm}
\label{fig:area_comp}
\end{figure}

\begin{figure}[t]
\centering
\includegraphics[width=\linewidth]{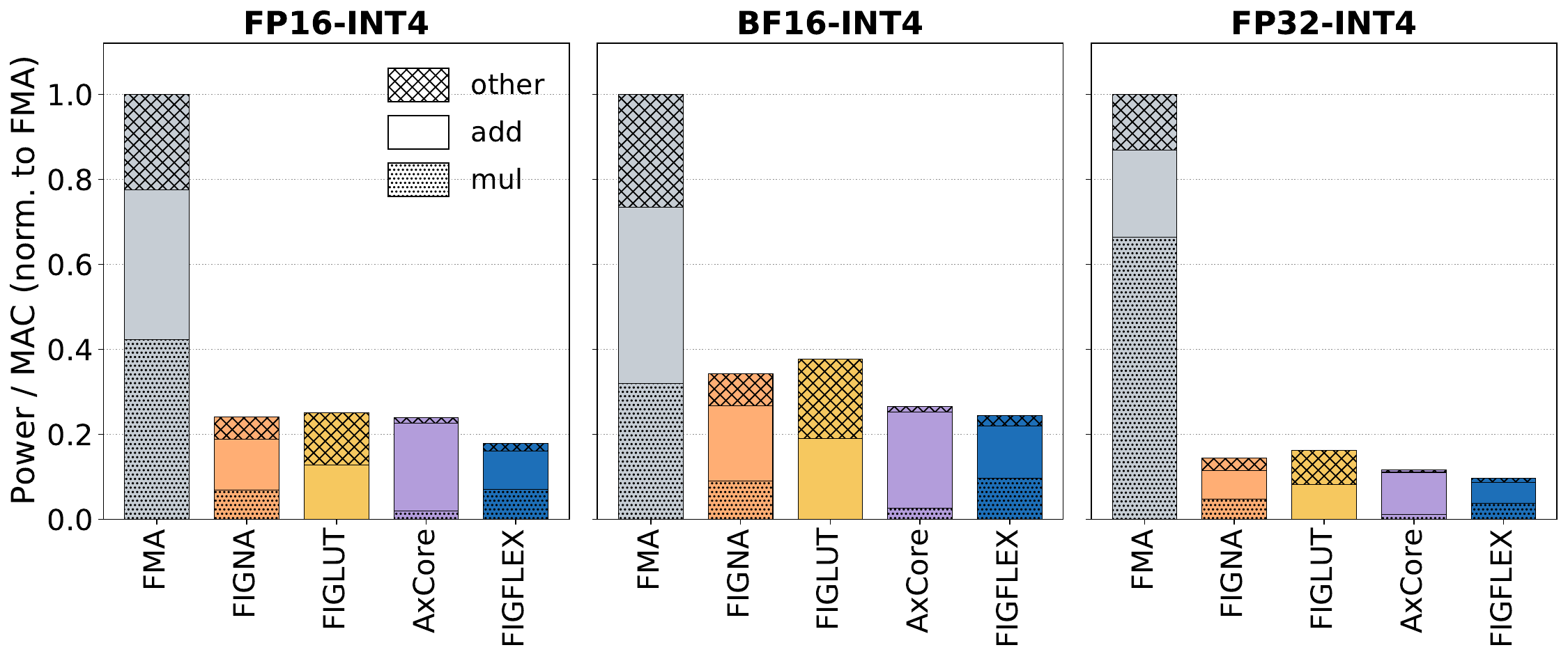}
\vspace{-7mm}
\caption{\small Accelerator power breakdown (excluding SRAM) across the compared engines
under the common iso-frequency budget.}
\vspace{-3mm}
\label{fig:power_comp}
\end{figure}

\section{Related Work}
\label{sec:related}

\subsection{Precision-Scalable DNN Accelerators}
\emph{Bit-serial} designs (Stripes~\cite{stripes}, Bit-Pragmatic~\cite{bitpragmatic}, UNPU~\cite{unpu}) process one bit-plane per cycle, so cost tracks precision but latency grows with it. \emph{Bit-parallel composable} designs build wide operations from narrow ones: BitFusion~\cite{bitfusion} fuses 2-bit multipliers, BitBlade~\cite{bitblade} amortizes their shifters across the array, and FlexiBit~\cite{flexibit} generalizes this to arbitrary formats. \accelerator is bit-parallel and, like BitBlade, hoists the shift networks out of the PE, but it scales only the weight operand, allocates by radix-4 digit rather than by bit, and runs the intra- and inter-weight reduction on one integer adder tree after block-floating-point pre-alignment.

\subsection{Beyond Weight-Only Quantization}
Activation quantization~\cite{llmint8,ant,olive,tender,atom,qserve} and KV-cache quantization~\cite{kivi,kvquant,gear} attack the FP$\times$FP attention operators and the memory-bound decode phase; they are complementary to the FP-INT linear GEMM that FIGFLEX accelerates. Codebook quantizers (SqueezeLLM~\cite{squeezellm}, QuIP\#~\cite{quipsharp}, AQLM~\cite{aqlm}, VPTQ~\cite{vptq}, QTIP~\cite{qtip}) and sparse-outlier formats (SpQR~\cite{spqr}, OWQ~\cite{owq}) reach strong quality at 2--3 bits but decode through a codebook or an index, which reintroduces a dequantization step or needs a LUT datapath.

\subsection{Algorithm--Hardware Co-Design for Quantized LLMs}
Mixed-precision PTQ (SliM-LLM~\cite{slimllm}, AMQ~\cite{amq}, LLM-MQ~\cite{llmmq}) selects bit-widths for model quality alone, and precision-flexible accelerators~\cite{figlut,bitblade,flexibit} are evaluated on externally produced models; Tender~\cite{tender} adapts the numeric format to the datapath but not the per-group allocation. \accelerator closes this loop: the allocator's precision set $\{$W4, W3, W2, W1$\}$ is exactly the set the array executes at $32/b$ MAC/cycle.

\section{Conclusion}

We presented \accelerator, an algorithm--hardware co-design that makes low-bit weight quantization pay off in computation as well as in memory. Its radix-4 FP-INT processing element is bit-parallel yet precision-scalable: every lane resolves two weight bits, so INT4 to INT1 weights run at a rate inversely proportional to their bit-width, and the datapath adds no error beyond shared-exponent alignment. With a compensation-aware mixed-precision quantizer, \accelerator attains 5.85 and 8.77 WikiText-2 perplexity at 3- and 2-bit average precision on LLaMA-2-7B, the lowest among the compared affine post-training quantizers. On the linear GEMM layers it delivers $1.56\times$ and $1.42\times$ the whole-chip throughput per area of AxCore~\cite{axcore} and FIGLUT~\cite{figlut} at 3 bits, which a first-order request model translates into 2.2--2.9$\times$ (F3) and 2.9--3.7$\times$ (F2) shorter batched requests than GPU low-bit kernels at the same bit-width and compute area.

\bibliographystyle{IEEEtranS}
\bibliography{refs}

\end{document}